\documentclass[12pt]{article}

\usepackage[cp1251]{inputenc}
\usepackage[T1,T2A]{fontenc}
\usepackage[russian]{babel}

\usepackage{amsmath,amssymb,bm}
\usepackage[a4paper,margin=2cm,right=3cm,includefoot,nohead,nomarginpar]{geometry}

\usepackage{dsmacro,cite}

\usepackage{graphicx}

\def\Im{\mathop{\mathrm{Im}}\nolimits}
\def\const{\mathop{\mathrm{const}}\nolimits}
\def\sign{\mathop{\mathrm{sign}}\nolimits}

\begin{document}

\title{Эволюция спектра и переход металл"--~изолятор в локальных приближениях для многоэлектронных моделей}

\author{А.О. Анохин, А.В. Зарубин, В.Ю. Ирхин}



\maketitle

\begin{abstract}

В рамках многоэлектронной $s$"~$d$~обменной модели Шубина"--~Вонсовского и модели Хаббарда выводятся самосогласованные уравнения для одночастичной запаздывающей функции Грина в представлении многоэлектронных $X$-операторов Хаббарда методом расцепления цепочки уравнений движения. Анализируется общая структура одноузельных приближений и их связь с приближением когерентного потенциала (ПКП, CPA) и теорией динамического эффективного поля (DMFT).

Используя самосогласованные приближения, мы детально исследуем картину эволюции электронного спектра при изменении затравочных параметров модели (константы связи, концентрации носителей тока). Обсуждается влияние различных факторов (кондовское многоэлетронное рассеяние, размытие, обусловленное затуханием, собственная динамика подсистемы локализованных моментов) на форму плотности состояний $N(E)$ с учетом взаимодействия.

Показано, что использование локаторного представления позволяет в ряде случаев избежать неаналитичностей в приближенных выражениях для функции Грина. Этот подход позволяет воспроизвести при определенных значениях параметров трехпиковую структуру $N(E)$  вблизи перехода металл"--~изолятор.

\end{abstract}

\section{Введение}

Одним из наиболее важных и интересных направлений теоретической и экспериментальной физики конденсированного состояния вещества, активно развивающимся в последнее время, является изучение многоэлектронных эффектов в соединениях на основе переходных и редкоземельных элементов с $3d$, $4f$ и~$5f$ незаполненными оболочками. Это обусловлено громадным разнообразием физических явлений, наблюдающихся в данных системах и их практической значимостью. К числу таких физических явлений относятся широко известный переход металл"--~изолятор (моттовский переход), формирование состояний с большой массой носителей заряда в тяжелофермионных системах, где происходят различные фазовые переходы с участием сверхпроводящей, магнитоупорядоченной, диэлектрической и металлической фаз, а также переходы между состояниями с локализованными и делокализованными магнитными моментами. Сюда же относятся явление высокотемпературной сверхпроводимости (с рядом известных оговорок), нефононные механизмы формирования сверхпроводящего состояния, валентные флуктуации,  гигантское магнетосопротивление, аномальные электронные свойства и магнетизм решеток Кондо. Как выяснилось при дальнейшей разработке теории, однопримесный эффект Кондо играет очень важную роль и в правильном понимании и описании моттовского перехода.

Характерной особенностью этих физических систем является значительный вклад многоэлектронных эффектов в наблюдаемые свойства, обусловленный сильным кулоновским взаимодействием между электронами незаполненных внутренних оболочек  на узле и их взаимодействием с коллективизированными электронами внешних оболочек. При этом величина внутриатомного кулоновского взаимодействия сравнима или больше ширины затравочной зоны (характеризуется величиной интеграла перекрытия, переноса). Такие системы принято называть сильно коррелированными системами или системами с сильным межэлектронным взаимодействием.

Начиная с первых попыток теоретического описания Хаббардом~\cite{Hubbard-I:1963,Hubbard-III:1964} перехода металл"--~изолятор и введения им многоэлектронных операторов, для  описания явлений в сильнокоррелированных системах применялся весь спектр теоретических методов. Широко использовались методы основанные на приближенном расцеплении цепочки уравнений движения для функций Грина, метод континуального интегрирования (спин"=флуктуационные теории), разнообразные расширения теории эффективного поля, метод вспомогательных бозонов и фермионов в континуальном интеграле по когерентным состояниям, нестандартные диаграммные техники, различные методы численного моделирования и прочее.

Сложность построения согласованного теоретического описания этих систем в рамках модельных подходов обусловлена тем, что в таких системах происходит перестройка основного состояния вызванная сильными электронными корреляциями и, как следствие, при промежуточных значениях параметра взаимодействия (а именно этот случай и оказывается наиболее интересным) имеются одновременно два существенно различные типа элементарных возбуждений~"--- фермижидкостного типа и "<хаббардовские"> (связанные с локализованными состояниями). При этом оба типа возбуждений как правило связаны с одной и той же системой электронных состояний и последовательное описание в рамках теории возмущений оказывается не только недостаточным, но трудно осуществимым. Отсюда возникает необходимость построения различных интерполяционных схем, являющихся дополнительными по отношению друг к другу с точки зрения корректно описываемых ими свойств.

Наметившийся в последние пару десятилетий прогресс в построении единого описания во всей области параметров модельных гамильтонианов сильно коррелированных систем стал возможен благодаря замечательному наблюдению, сделанному Фоллхардтом и Метцнером~\cite{PRL.62.324}. В рамках исследований модели Хаббарда они обнаружили, что в пределе больших размерностей пространства~$d$ или большого решеточного координационного числа~$z$ и при подходящей перенормировке затравочных интегралов переноса законом сохранения квазиимпульса в вершинах взаимодействия можно пренебречь, так что вся физика задачи определяется исключительно динамикой на одном узле. В этом случае решение задачи сводится к нахождению динамики узла во внешнем динамическом поле произвольного вида (содержащем как запаздывающие, так и опережающие компоненты), которое в дальнейшем находится из некоторого условия самосогласования. Вообще говоря, расчет одночастичной функции Грина для такой задачи  формально не проще, чем её нахождение для исходной решеточной задачи, так как с точки зрения теории возмущений слабой связи необходимо учитывать всё тот же полный набор скелетных диаграмм в разложении собственно-энергетической части одночастичной функции Грина: хотя зависимости от квазиимпульса нет, остающаяся временная (энергетическая) зависимость диаграмм нетривиальна.

Этот модельно не специфический факт и лег в основу разработанного авторами обзора~\cite{RMP.68.13} метода "<динамического эффективного поля"> (Dynamical Mean Field Theory, DMFT), которые переформулировали задачу о примесном узле в поле в виде задачи о примеси взаимодействующей простым образом с термостатом свободных фермионов с самосогласованным определением его параметров. В зависимости от использованного для внешнего поля представления, такая задача сводится к решению однопримесной модели Андерсона (используется спектральное представление для поля виде интегральной суммы) или, гораздо реже, однопримесной модели Вольфа~\cite{PR:Wolff:124} (используется спектральное представление для поля в виде непрерывной дроби типа Якоби-Чебышева) с самосогласованным определением параметров гамильтониана моделей.

Такая переформулировка~"--- сведение нульмерной задачи к эффективной (квази)одно\-мерной,~"--- позволяет применять к исследованию сильнокоррелированных систем и описывающих их моделей весь спектр теоретических и численных методов, разработанных для изучения одномерных систем (динамические и термодинамические свойства которых к настоящему времени изучены очень хорошо), а также применять наработанные в этой области интуиции и аналогии для понимания поведения дву- и трехмерных моделей сильно коррелированных систем.

Вместе с тем, указанное отображение на одномерную эффективную модель является скорее удобным средством, чем принципиальным моментом, когда речь идет о непосредственном вычислении спектров возбуждений системы и прочих свойств. В самом деле, различные методы решения эффективной примесной задачи, основанные на том или ином варианте численной ренормгруппы или численной диагонализации, используют подобное отображение, а появившийся не так давно и уже ставший очень популярным метод Монте-Карло с непрерывным временем (CT-QMC)~\cite{PRB:Rubtsov:72,PRL:Werner:97}~"--- нет. Более принципиальным кажется учет возможного \textit{целевого} основного состояния (например, антиферромагнитного) при построении нульмерной примесной модели (или семейства моделей в общем случае) и определении условий самосогласования на эффективное динамическое поле (поля). Целевое  основное состояние выбирается исходя из того круга физических явлений, которые необходимо описать.

В базовой  форме схема метода DMFT может быть изложена следующим образом.
\begin{itemize}
 \item Для заданного гамильтониана решеточной модели сильно коррелированной системы и учетом \textit{возможного} спонтанного нарушения симметрии в основном состоянии нужно выписать соответствующую эффективную однопримесную модель (или семейство моделей) и условие самосогласования на эффективное динамическое поле (поля). Обычная форма такого условия связывает диагональный матричный элемент одночастичной функции Грина исходной решеточной задачи с одночастичной функцией Грина примеси или непосредственно с динамическим полем эффективной примесной задачи.

 \item Найти необходимые корреляционные функции  эффективной однопримесной задачи для произвольного динамического поля. Как правило, ищется одночастичная функция Грина, поскольку именно она или связанная с ней величина фигурирует в типичных уравнениях на нахождение самосогласованного значения эффективного динамического поля.

\item Уравнения на самосогласованное динамическое поле и набор корреляционных функций, зависящих от этого поля, образуют замкнутую систему уравнений, которую и необходимо решить.
\end{itemize}
На практике изложенная общая схема обычно реализуется численно. Для этого необходимо знать уравнения для нахождения самосогласованного динамического поля и выражения для корреляционных функций примесной модели, в частности, для одночастичной функции Грина. Как правило уравнения для нахождения самосогласованного динамического поля известны в явной форме. Совершенно иным образом обстоит дело со знанием точной явной или неявной функциональной зависимости одночастичной функции Грина для однопримесной задачи от произвольного динамического поля. Точное знание такой зависимости является скорее исключением, чем правилом, и в аналитическом виде эта зависимость может быть установлена лишь приближенно. Хорошей альтернативой знанию явной или неявной зависимости функции Грина от поля является применение методов непосредственного численного моделирования квантовых систем, в которых необходимые корреляционные функции вычисляются непосредственно для заданного значения динамического поля.

Таким образом центральным моментом любой практической реализации схемы DMFT является построение эффективного способа решения примесной задачи в произвольном внешнем поле (так называемого \textit{примесного сольвера}) на основе точной или приближенной численной процедуры нахождения требуемых корреляционных функций с использованием хорошо известный численных методов типа точной диагонализации, численной ренормгруппы, метода Монте-Карло и~пр., либо на основе корректного с точки зрения описываемых свойств аналитического приближения для одночастичной функции Грина примесной задачи с последующей его численной реализацией. Если же учитывать локальный характер эффективной примесной модели, то в качестве приближенного сольвера возможно попытаться использовать какое-либо локальное приближенное решение для одночастичной функции Грина исходной решеточной модели.

Исследованию некоторых приближений такого типа~\cite{Irkhin:2001:MIT,Irkhin:1999:KEP,Anokhin:1991:OTM} и посвящена данная работа. Для $s$"~$d$~обменной модели Шубина"--~Вонсовского и модели Хаббарда выводятся самосогласованные уравнения для одночастичной функции Грина в представлении многоэлектронных операторов методом расцепления цепочки уравнений движения. Обсуждается общая структура одноузельных приближений в рассматриваемых многоэлектронных моделей с учетом возможного замороженного беспорядка типа замещения и их связь с приближением DMFT, а также с известным в теории неупорядоченных систем приближением когерентного потенциала (ПКП, CPA). Далее приводится детальная картина эволюции электронного спектра при изменении затравочных параметров модели (константы связи, концентрации носителей тока). Обсуждается влияние различных факторов (кондовское рассеяние, собственная динамика подсистемы локализованных моментов, затухание)  на форму плотности состояний системы.

\section{Гамильтонианы моделей и уравнения движения для одначастичной функции Грина}

В многоэлектронном представлении обобщенных проекционных операторов Хаббарда~\cite{Hubbard-IV:1965,Irkhin:1994:MEO,Izyumov:1995}
\begin{equation}
X_{i}^{\alpha \beta }=|i\alpha \rangle \langle i\beta |.
\label{eq:X}
\end{equation}
одночастичный оператор рождения электрона на узле~$i$ со спином $\sigma $ записывается как
\begin{equation}
c_{i\sigma }^{\dag }=X_{i}^{\sigma 0}+\sigma X_{i}^{2-\sigma }.
 \label{eq:c2X}
\end{equation}
В  этом представлении гамильтониан взаимодействия диагонален и модель Хаббарда имеет вид
\begin{equation}
\mathcal{H}=\sum_{\mathbf{k},\sigma }t_{\mathbf{k}}(X_{-\mathbf{k}}^{\sigma 0}+\sigma X_{-\mathbf{k}}^{2-\sigma })(X_{\mathbf{k}}^{0\sigma }+\sigma X_{\mathbf{k}}^{-\sigma 2})+U\sum_{i}X_{i}^{22},
 \label{eq:HHM:X:m}
\end{equation}
где~$X_{\mathbf{k}}^{\alpha \beta }$~"--- фурье-образ оператора Хаббарда, $t_{\mathbf{k}}$~"--- одночастичный закон дисперсии, $U$~"--- кулоновское отталкивание на узле. Для $s$"~$d$~обменной модели Шубина"--~Вонсовского гамильтониан в многоэлектронном представлении имеет похожую структуру
\begin{equation}
\mathcal{H}_{s-d}=\sum_{\mathbf{k},\sigma} t_{\mathbf{k}} (f^{\dag}_{\mathbf{k},\sigma} + \sigma g^{\dag}_{\mathbf{k},\sigma}) (f_{\mathbf{k},\sigma} + \sigma g_{\mathbf{k},\sigma}) +\mathcal{H}_{\text{int}},
 \label{eq:SDM:X:m}
\end{equation}
\begin{equation}
\mathcal{H}_{\text{int}}= - IS     \sum_{i}\sum_{\mu=-S-\frac{1}{2}}^{S+\frac{1}{2}}
                            \XO{i}{\mu,\,+}{\mu,\,+} +
                     I(S+1) \sum_{i}\sum_{\mu=-S+\frac{1}{2}}^{S-\frac{1}{2}}
                            \XO{i}{\mu,\,-}{\mu,\,-},
\label{eq.SD.interaction}
\end{equation}
где введено многоэлектронное представление для одночастичных операторов рождения и уничтожения
\begin{equation}
   c^{\dag}_{i\sigma} = f^{\dag}_{i\sigma} + \sigma g^{\dag}_{i\sigma}
\label{eq.c.SD.oper}
\end{equation}
\begin{equation}
   f^{\dag}_{i\sigma} = \sum_{M=-S}^{S}
                        \left[
                           \left(
                              \frac{S+\sigma M+1}{2S+1}
                           \right)^\frac{1}{2} \XO{i}{M+\sigma/2,+}{M,0} +
                           \left(
                             \frac{S+\sigma M}{2S+1}
                           \right)^\frac{1}{2} \XO{i}{M,2}{M-\sigma/2,-}
                        \right]
\label{eq.f.SD.oper}
\end{equation}
\begin{equation}
   g^{\dag}_{i\sigma} =
                        \sum_{M=-S}^{S}
                        \left[
                           \left(
                              \frac{S-\sigma M}{2S+1}
                           \right)^\frac{1}{2} \XO{i}{M+\sigma/2,-}{M,0} +
                           \left(
                             \frac{S-\sigma M+1}{2S+1}
                           \right)^\frac{1}{2} \XO{i}{M,2}{M-\sigma/2,+}
                        \right]
\label{eq.g.SD.oper}
\end{equation}
Здесь $I$~"--- константа $s$"~$d$~обменного взаимодействия, $S$~"--- величина локализованного спина. Отметим, что структуры многоэлектронных представлений для одноэлектронных операторов рождения и уничтожения для модели Хаббарда и $s$"~$d$~обменной модели аналогичны. В~данном случае $X$"~оператор~(\ref{eq:X}) имеет вид
\[
X_{i}^{M+\sigma /2,\alpha ;M}=|iM+\sigma /2,\alpha \rangle \langle iM|,
\]
где~$|iM\rangle $~"--- состояние без электронов проводимости и~с~проекцией локализованного спина~$M$, и~$|m,\alpha \rangle $~"--- однократно занятое состояние с~полным спином на~узле~$S+\alpha /2$ и~его проекцией~$m$~(которые остаются при~$|I|\rightarrow \alpha \infty $), $\sigma $~"--- проекция спина.

Для вычисление одночастичной плотности состояний рассмотрим запаздывающую антикоммутаторную функцию Грина
\begin{equation}
G_{\mathbf{k}\sigma }(E)=\langle \!\langle c_{\mathbf{k}\sigma }|c_{\mathbf{k}\sigma }^{\dagger }\rangle \!\rangle _{E},\quad \Im E>0
 \label{eq:HM:GF}
\end{equation}
и для плотности одночастичных состояний (Density of State, DOS) имеем стандартное выражение
\begin{equation}
   N_{\sigma }(E) = -\frac{1}{\pi } \Im \sum_{\mathbf{k}} G_{\mathbf{k}\sigma }(E+i0).
\end{equation}
Для вычисления $G_{\mathbf{k}\sigma }(E)$ строится цепочка уравнений движения
\begin{equation}
E\GKets{\mathcal{A}}{c^{\dag }_{\mathbf{k}\sigma }}{E}
=\langle \lbrack \mathcal{A},c_{\mathbf{k}\sigma }^{\dagger }]_{+}\rangle
+\langle \!\langle \lbrack \mathcal{A},\mathcal{H}]|c_{\mathbf{k}\sigma }^{\dagger }\rangle \!\rangle _{E}
\label{eq:HM:EM}
\end{equation}
для парциальных функций Грина в многоэлектронном представлении
\begin{equation}
G_{\mathbf{k}\sigma }(E)=\langle \!\langle X_{\mathbf{k}}^{0\sigma }|c_{\mathbf{k}\sigma }^{\dagger }\rangle \!\rangle _{E}+\sigma \langle \!\langle X_{\mathbf{k}}^{-\sigma 2}|c_{\mathbf{k}\sigma }^{\dagger }\rangle \!\rangle _{E},
\quad \Im E>0
\end{equation}
для модели Хаббарда, и
\begin{equation}
   G_{\mathbf{k}\sigma }(E) =
                  \GKets{f_{\mathbf{k}\sigma }}{c^{\dag}_{\mathbf{k}\sigma }}{E} +
            \sigma\GKets{g_{\mathbf{k}\sigma }}{c^{\dag}_{\mathbf{k}\sigma }}{E} ,
\quad \Im E>0
\label{eq.el.Green.SD}
\end{equation}
для $s$"~$d$~обменной модели соответственно.
Для получения явных выражений для функции Грина требуется выразить функции Грина высших порядков через функции низшего порядка, что обычно делается с помощью той или иной процедуры расцепления цепочки уравнения движения.

\section{Процедура расцепления и несамосогласованные выражения для функции Грина}

Для получения уравнений на определение функции Грина цепочка уравнений движения выписывалась до второго порядка по флуктуациям бозе-полей (спинового и зарядового) и далее, после проведения процедуры симметризации необходимой здесь для явного учета частично-дырочной симметрии, расцеплялась в этом порядке
(что грубо соответствует первому порядку в разложении по обратному координационному числу $1/z$).
Отметим, что полученные таким образом приближения для функции Грина, вообще говоря, не носят локального характера
и не определяют функцию Грина самосогласованным образом.

При половинном заполнении  одночастичная функция Грина в модели Хаббарда
может быть записана в локаторном представлении~\cite{Irkhin:2001:MIT}.

\begin{equation}
G_{\mathbf{k}\sigma }(E)=[F_{\mathbf{k}\sigma
}(E)-t_{\mathbf{k}}]^{-1},\quad F_{\mathbf{k}\sigma
}(E)=\frac{b_{\mathbf{k}\sigma }(E)}{a_{\mathbf{k}\sigma }(E)},
\label{eq:HM:GF:1}
\end{equation}
где $F_{\mathbf{k}\sigma}$~"--- обратный локатор, который не зависит от квазиимпульса в локальных приближениях,
\begin{eqnarray}
a_{\mathbf{k}\sigma }(E) &=&1+\frac{U^{2}}{E^{2}}\sum_{\mathbf{q}}\chi _{\mathbf{k-q}}t_{\mathbf{q}}G_{\mathbf{q}}^{0}(E)+\frac{2U}{E}\sum_{\mathbf{q}}t_{\mathbf{q}}n_{\mathbf{q}}G_{\mathbf{q}}^{0}(E),
\label{eq:HM:A:1} \\
b_{\mathbf{k}\sigma }(E) &=&F^{0}(E)+\frac{2U}{E}\sum_{\mathbf{q}}t_{\mathbf{q}}^{2}n_{\mathbf{q}}G_{\mathbf{q}}^{0}(E).
\label{eq:HM:B:1}
\end{eqnarray}
Функция Грина в приближении "<Хаббард-I">~\cite{Hubbard-I:1963} есть
\[
G_{\mathbf{k}}^{0}(E)=[F^{0}(E)-t_{\mathbf{k}}]^{-1},\quad F^{0}(E)=E-\frac{U^{2}}{4E},
\]
а функция $F^{0}(E)$ есть соответствующий обратный локатор.

Отметим, что фермиевские функции $n_{\mathbf{q}\sigma }=\langle c^{\dag}_{\mathbf{q}\sigma } c_{\mathbf{q}\sigma }\rangle $ в выражении для обратного локатора соответствуют учету вкладов кондовского типам (многоэлектроннх процессов рассеяния).

Для  $s$"~$d$~обменной модели в квазиклассическом пределе $S\to \infty $, $IS\to \const $ можно поучить ряд аналогичных уравнений~\cite{Anokhin:1991:OTM}. Запаздывающая одночастичная функция Грина в несамосогласованном случае имеет вид~(\ref{eq:HM:GF:1}), где нужно подставить обратный локатор в приближении "<Хаббард-I">
\begin{equation}
 b(E)=y(E)=E - \frac{(IS)^2}{4E}
 \label{eq:b:sd}
\end{equation}
и положить
\begin{equation}
 a_{\mathbf{k}}(E)=1+{\left(\frac{IS}{E}\right)}^2
         \sum_{\mathbf{q}} \frac{\chi_{\mathbf{k-q}}}{(2S)^2}
         \frac{t_\mathbf{q}}{y(E)-t_{\mathbf{q}}},
 \label{eq:a:sd}
\end{equation}
$\chi_{\mathbf{q}}=\langle \mathbf{S}_{\mathbf{q} \text{tot}}\mathbf{S}_{\mathbf{-q} \text{tot}} \rangle$~"--- полная магнитная восприимчивость. В формулах (\ref{eq:b:sd}), (\ref{eq:a:sd}) отсутствуют вклады с фермиевскими функциями, так как в квазиклассическом случае они формально малы по $1/S$. Однако, если превысить точность, они могут быть восстановлены, что приведет к структуре, аналогичной таковой для модели Хаббарда~(\ref{eq:HM:A:1}) и~(\ref{eq:HM:B:1}).

Для модели Хаббарда с $U=\infty $, используя антиперестановочность фермиевских операторов и~обозначая~$\varepsilon _{\mathbf{k}}=-t_{\mathbf{k}}$, получим гамильтониан в~представлении $X$"~операторов
\[
\mathcal{H}=\sum_{\mathbf{k},\sigma }\varepsilon _{\mathbf{k}}X_{-\mathbf{k}}^{0\sigma }X_{\mathbf{k}}^{\sigma 0}.
\]
Далее мы~будем вычислять одночастичную  функцию Грина в~многоэлектронном представлении
\[
\widetilde{G}_{\mathbf{k}\sigma }(E)=\langle \!\langle X_{\mathbf{k}}^{\sigma 0}|X_{-\mathbf{k}}^{0\sigma }\rangle \!\rangle _{E},\quad \Im E>0,
\]
решая цепочку уравнений движения.
Приближение типа~"<Хаббард"~I">, даёт
\begin{equation}
\widetilde{G}_{\mathbf{k}\sigma }^{0}(E)=[F_{\sigma }^{0}(E)-\varepsilon _{\mathbf{k}}]^{-1},\quad F_{\sigma }^{0}(E)=\frac{E}{n_{0}+n_{\sigma }},
\label{eq:HMUI:GF:0}
\end{equation}
где~$F_{\sigma }^{0}(E)$~"--- обратный локатор и
$\varepsilon _{\mathbf{k}\sigma }=\varepsilon _{\mathbf{k}}(n_{0}+n_{\sigma })$~"--- энергетический спектр в~этом приближении,
$n_{\alpha }=\langle X_{i}^{\alpha \alpha }\rangle $~"--- числа заполнения для дырок и однократно заполненных состояний ($\alpha =0,\ \sigma $),
\[
\chi _{\mathbf{q}}^{\sigma -\sigma }=\langle S_{\mathbf{q}}^{\sigma }S_{-\mathbf{q}}^{-\sigma }\rangle =\langle X_{\mathbf{q}}^{\sigma -\sigma }X_{-\mathbf{q}}^{-\sigma \sigma }\rangle ,
\]
--- поперечная корреляционная функция спиновых операторов,
\[
\lambda _{\mathbf{q}}^{\sigma }=\langle \delta X_{\mathbf{q}}^{\sigma \sigma }\delta X_{-\mathbf{q}}^{\sigma \sigma }\rangle =\langle \delta (X_{\mathbf{q}}^{00}+X_{\mathbf{q}}^{-\sigma -\sigma })\delta (X_{-\mathbf{q}}^{00}+X_{-\mathbf{q}}^{-\sigma -\sigma })\rangle ,
\]
--- продольная корреляционная функция спиновых флуктуаций,
\begin{equation}
n_{\mathbf{k}\sigma }=\langle X_{-\mathbf{k}}^{0\sigma }X_{\mathbf{k}}^{\sigma 0}\rangle =-\frac{1}{\pi }\Im \int\limits_{-\infty }^{+\infty }\widetilde{G}_{\mathbf{k}\sigma }(E)f(E)\,dE,
\label{eq:HMUI:nk}
\end{equation}
--- одночастичная функция распределения, которая находится из~спектрального представления.

При расцеплении цепочки уравнения движения во втором порядке по флуктуациям $X$-операторов функция Грина в~локаторной форме имеет вид
\begin{equation}
\widetilde{G}_{\mathbf{k}\sigma }(E)=[F_{\mathbf{k}\sigma }(E)-\varepsilon _{\mathbf{k}}]^{-1},\quad F_{\mathbf{k}\sigma }(E)=\frac{b_{\mathbf{k}\sigma }(E)}{a_{\mathbf{k}\sigma }(E)},
\label{eq:HMUI:GF:1}
\end{equation}
где
\begin{eqnarray}
a_{\mathbf{k}\sigma }(E) &=&n_{0}+n_{\sigma }+
 \nonumber \\
&&+\sum_{\mathbf{q}}\varepsilon _{\mathbf{k-q}}\frac{\chi _{\mathbf{q}}^{\sigma -\sigma }
       +n_{\mathbf{k-q}-\sigma}}{E-\varepsilon_{\mathbf{k-q}\sigma}}+
 \nonumber \\
&&+\sum_{\mathbf{q}}\varepsilon_{\mathbf{k-q}}
            \frac{\lambda _{\mathbf{q}}^{-\sigma }}{E-\varepsilon_{\mathbf{k-q}\sigma }},
 \label{eq:HMUI:GF:01:a} \\
b_{\mathbf{k}\sigma }(E) &=&E+\sum_{\mathbf{q}}\varepsilon _{\mathbf{k-q}}^{2}
\frac{n_{\mathbf{k-q}-\sigma }}{E-\varepsilon_{\mathbf{k-q}-\sigma }}.
 \label{eq:HMUI:GF:01:b}
\end{eqnarray}

Для $s$"~$d$~обменной модели можно получить аналогичные выражения. Здесь мы приведем только выражения для локатора одночастичной функции Грина для частного случая $I\rightarrow \pm \infty $. Вместо введенных выше операторов~(\ref{eq.f.SD.oper}) и~(\ref{eq.g.SD.oper}) в этом случае имеем
\begin{eqnarray*}
g_{i\sigma +}^{\dagger } &=&\sum_M\{(S+\sigma M+1)/(2S+1)\}^{1/2} X_i(M+\frac \sigma 2,+;M), \\
g_{i\sigma -}^{\dagger } &=&\sum_M\sigma \{(S-\sigma M)/(2S+1)\}^{1/2} X_i(M+\frac \sigma 2,-;M),
\end{eqnarray*}
Тогда гамильтониан приобретает вид
\begin{equation}
\mathcal{H}=\sum_{\mathbf{k}\sigma }t_{\mathbf{k}} g_{\mathbf{k}\sigma \alpha }^{\dagger }g_{\mathbf{k}\sigma \alpha } +\mathcal{H}_d,\quad \alpha =\sign I.
\label{eq:sd:H}
\end{equation}
Функция Грина может быть записана в виде
\begin{equation}
G_{\mathbf{k}\sigma \alpha }(E)=\langle \!\langle g_{\mathbf{k}\sigma \alpha }|g_{\mathbf{k}\sigma \alpha }^{\dagger }\rangle \!\rangle _E =
[F_{\mathbf{k}\sigma \alpha }(E)-t_{\mathbf{k}}]^{-1},\quad
F_{\mathbf{k}\sigma \alpha }(E)=\frac{b_{\mathbf{k}\sigma \alpha }(E)}{a_{\mathbf{k}\sigma \alpha }(E)}.
\label{eq:sd:EGF:0}
\end{equation}
Выражение для несамосогласованного локатора имеет ту же форму, что и в  модели Хаббарда, но величины $a_{\mathbf{k}\sigma \alpha }(E)$ и $b_{\mathbf{k}\sigma \alpha }(E)$ записываются как
(ср. \cite{Irkhin:1999:KEP})
\begin{eqnarray}
a_{\mathbf{k}\sigma \alpha }(E) &=&P_{\sigma \alpha }+\sum_{\mathbf{q}} \frac{t_{\mathbf{k-q}}}{(2S+1)^2} \left[ \frac{\chi _{\mathbf{q}}^{\sigma -\sigma }+(2S+1)n_{\mathbf{k-q}-\sigma \alpha }} {E-t_{\mathbf{k-q}-\sigma \alpha }} +\frac{\chi _{\mathbf{q}}^{zz}}{E-t_{\mathbf{k-q}\sigma \alpha }}\right] ,
\label{eq:sd:a0} \\
b_{\mathbf{k}\sigma \alpha }(E) &=&E-\sum_{\mathbf{q}} \frac{t_{\mathbf{k-q}}^2}{2S+1}\frac{n_{\mathbf{k-q}-\sigma \alpha }} {E-t_{\mathbf{k-q}-\sigma \alpha }},
\label{eq:sd:b0}
\end{eqnarray}
где
\[
t_{\mathbf{k}\sigma \alpha }=P_{\sigma \alpha }t_{\mathbf{k}},\quad
P_{\sigma \alpha }=\frac{\widetilde{S}+1/2+\alpha \sigma \langle S_{\text{tot}}^z\rangle -\alpha n/2}{2S+1},\quad
\widetilde{S}=S+\frac{\alpha }{2},
\]
$\chi _{\mathbf{q}}^{\sigma -\sigma }$~"--- поперечная корреляционная функция спиновых операторов, $\chi _{\mathbf{q}}^{zz}$~"--- продольная корреляционная функция спиновых операторов, $n_{\mathbf{k}\sigma \alpha }$~"--- одночастичная функция распределения
\[
\chi _{\mathbf{q}}^{\sigma -\sigma }=\langle S_{\text{tot}\mathbf{-q}}^\sigma S_{\text{tot}\mathbf{q}}^{-\sigma }\rangle ,\quad
\chi _{\mathbf{q}}^{zz}=\langle S_{\mathrm{tot}\mathbf{-q}}^{z}S_{\mathrm{tot}\mathbf{q}}^{z}\rangle ,\quad
n_{\mathbf{k}\sigma \alpha }=\langle g_{\mathbf{k}\sigma \alpha }^{\dagger } g_{\mathbf{k}\sigma \alpha }\rangle .
\]
Здесь $S_{\mathrm{tot}}^{\sigma }$, $S_{\mathrm{tot}}^{z}$~"--- операторы полного спина (включая вклады пустых и~однократно занятых состояний (см.~\cite{Anokhin:1991:OTM})):
\begin{eqnarray*}
S_{\mathrm{tot}}^{\sigma }
&=&\sum\limits_{M=-S}^{S}\sqrt{(S+\sigma M+1)(S-\sigma
M)}X_{i}^{M;M+\sigma }+
\\
&&+\sum\limits_{M=-\widetilde{S}}^{\widetilde{S}}
\sqrt{(\widetilde{S}+\sigma M+1)(\widetilde{S}-\sigma
M)}X_{i}^{M;M+\sigma },\\
S_{\mathrm{tot}}^{z} &=&\sum\limits_{M=-S}^{S}MX_{i}^{M;M}
+\sum\limits_{M=-\widetilde{S}}^{\widetilde{S}}MX_{i}^{M;M},
\end{eqnarray*}
причем  $\langle S_{\mathrm{tot}}^{z}\rangle $~"--- полная средняя намагниченность.

Отметим, что в приближении "<Хаббард"~I"> для $s$"~$d$ обменной модели $n_{\mathbf{k}\sigma \alpha }=P_{\sigma \alpha } f(t_{\mathbf{k}\sigma \alpha })$, где~$f(E)$~"--- функция Ферми. Вводя
\begin{equation}
G_{\mathbf{k}\sigma \alpha }^{0}(E)=[F_{\sigma \alpha
}^{0}(E)-t_{\mathbf{k}}]^{-1},\qquad F_{\sigma \alpha
}^{0}(E)=\frac{E}{P_{\sigma \alpha }},
\label{eq:sd:EGF:H1P}
\end{equation}
получаем
\begin{eqnarray}
a_{\mathbf{k}\sigma \alpha }(E) &=&P_{\sigma \alpha }+\sum_{\mathbf{q}}\frac{t_{\mathbf{k-q}}}{(2S+1)^{2}}\frac{\chi _{\mathbf{q}}^{\sigma -\sigma }+(2S+1)n_{\mathbf{k-q}-\sigma \alpha }}{P_{-\sigma \alpha }}G_{\mathbf{k-q}-\sigma \alpha }^{0}(E)+
 \nonumber \\
&&+\sum_{\mathbf{q}}\frac{t_{\mathbf{k-q}}}{(2S+1)^{2}}\frac{\chi _{\mathbf{q}}^{zz}}{P_{\sigma \alpha }}G_{\mathbf{k-q}\sigma \alpha }^{0}(E),
 \label{eq:sd:a0:loc} \\
b_{\mathbf{k}\sigma \alpha }(E) &=&E-\sum_{\mathbf{q}}\frac{t_{\mathbf{k-q}}^{2}}{2S+1}\frac{n_{\mathbf{k-q}-\sigma \alpha }}{P_{-\sigma \alpha }}G_{\mathbf{k-q}-\sigma \alpha }^{0}.
 \label{eq:sd:b0:loc}
\end{eqnarray}

Уравнение для~нахождения химического потенциала
\[
\sum_{\mathbf{k}}\langle g_{\mathbf{k}\sigma \alpha }^{\dagger } g_{\mathbf{k}\sigma \alpha }\rangle =\frac n2.
\]
позволяет при~заданной концентрации определять энергию Ферми носителей тока.

Между моделью Хаббарда с $U=\infty $ и $s$"~$d$~моделью при $S=1/2$, $I=-\infty $ имеется связь. Для этого случая явные выражения для операторов $g^{\dag}_{i\sigma -}$ имеют вид
\begin{equation}
  g^{\dag}_{i\sigma -} = \frac{\sigma}{\sqrt{2}} X_{i}(0,-;-\frac{\sigma}{2})
\end{equation}
Подставляя эти выражения в гамильтониан (\ref{eq:sd:H}) получаем
\begin{equation}
 H_{s-d}=\frac{1}{2}\sum_{\mathbf{k} \sigma}
           t_{\mathbf{k}}
           X_{\mathbf{k}} (0,-;-\frac{\sigma}{2})
           X_{\mathbf{-k}}(-\frac{\sigma}{2};0,-)
\end{equation}
Отсюда видно, что при заменах
\begin{equation}
t_{\mathbf{k}}/2\rightarrow t_{\mathbf{k}},\quad 2n_{\mathbf{k}}\rightarrow n_{\mathbf{k}},\quad n\rightarrow \delta
\label{eq:HM2sd}
\end{equation}
($\delta $~"--- концентрация "<дырок">), выражения (\ref{eq:sd:a0:loc}) и~(\ref{eq:sd:b0:loc}) для приближенной функции Грина в $s$"~$d$~модели переходят в выражения для модели Хаббарда. Поэтому результаты для модели Хаббарда будут совпадать с таковыми для $s$"~$d$ модели с $s=1/2$ и $I=-\infty $ и в дальнейшем мы можем обсуждать результаты только для $s$"~$d$ модели.

\section{Общая структура одноузельных приближений и связь с~DMFT}
\label{sec:dmft}

Выясним теперь общую структуру одноузельных приближений для обсуждаемых моделей. Для того, чтобы установить (скорее, подчеркнуть) связь одноузельных приближений с приближением когерентного потенциала, мы введем в гамильтонианы модели Хаббарда и $s$"~$d$~обменной модели одночастичный потенциал рассеяния, описывающий беспорядок
\begin{equation}
 H_{\text{dis}}=\sum_{i,\sigma} \xi_{i}c^{\dag}_{i\sigma}c_{i\sigma}
\end{equation}
где $\xi_{i}$~"--- случайная переменная на узле $i$ с некоторым распределением $P(\xi_{i})$, которая моделирует диагональный замороженный беспорядок типа замещения.

Межэлектронное взаимодействие в модели Хаббарда и в $s$"~$f$~обменной модели носит локальный характер, т.~е. диагонально в узельном представлении. Обозначим это взаимодействие $H_{i}[0]$. Воспользовавшись рассуждением~\cite{Shiba:1971}, можно построить систему самосогласовнных уравнений для запаздывающей (тябликовской) или мацубаровской функции Грина в одноузельном приближении для собственно-энергетической части.

Уравнение Дайсона для одночастичной функции Грина для  модели Хаббарда и $s$"~$f$~обменной модели имеет вид
\begin{equation}
G_{ij}(z)=G^{0}_{ij}(z) + \sum_{lm}G^{0}_{il}(z)\Sigma_{lm}(z)G_{mj}(z)
\end{equation}
где $\Sigma_{lm}(z)$~"--- полная собственно-энергетическая часть. Одноузельное приближение заключается в аппроксимации $\Sigma_{lm}\to\Sigma\delta_{lm}$ узельно-диагональной величиной. Тогда
\begin{equation}
G_{ij}(z)=G^{0}_{ij}(z) + \sum_{l}G^{0}_{il}(z)\Sigma(z)G_{lj}(z)
\label{eq:00}
\end{equation}
и $\Sigma$ находится из условия совпадения диагонального элемента функции Грина $G_{ii}(z)$ с диагональным элементом функции Грина эффективной примесной задачи на узле примеси. Эта эффективная примесная задача представляет собой задачу о движении электронов на решетке, в которой на всех узлах, кроме $i$-того, действует самосогласованное поле $\Sigma$, а на $i$-том узле действует многочастичный локальный потенциал рассеяния.

Далее везде для простоты полагаем все узлы исходной решетки магнитно эквивалентными, т.~е. рассматриваем только парамагнитное или ферромагнитное состояние.

Система уравнений для функции Грина в одноузельном приближении имеет следующую структуру. Из (\ref{eq:00}) для $G(z)$ получаем представление
\begin{equation}
G(z)=\int\frac{N_{0}(\epsilon)d\epsilon}{z-\epsilon -\Sigma(z)}
\label{eq:1}
\end{equation}
и для эффективной примесной задачи
\begin{equation}
G^{(i)}[\xi_{i}](z)=G_{0}^{(i)}(z) + G_{0}^{(i)}(z)V_{i}^{\text{eff}}[\xi_{i}](z)G^{(i)}[\xi_{i}](z)
\label{eq:2}
\end{equation}

Отметим, что это уравнение может быть интерпретировано как уравнение Дайсона для $0$-мерной системы, в которой имеется всего один узел, где происходит движение фермионов со свободным пропагатором $G_{0}^{(i)}(z)$ и с многочастичным локальным потенциалом рассеяния. В свою очередь $G_{0}^{(i)}(z)$ определена уравнением
\begin{equation}
G_{0}^{(i)}(z)=G(z) - G(z)\Sigma(z)G_{0}^{(i)}(z)
\label{eq:3}
\end{equation}
До сих пор считавшееся произвольным значение собственно-энергетической части $\Sigma(z)$ фиксируется условием самосогласования
\begin{equation}
G(z)={\langle G^{(i)}[\xi](z) \rangle}_{\xi}
\label{eq:4}
\end{equation}
Здесь для простоты записи мы не выписываем  зависимость Функций Грина от спиновых индексов, но явно указана возможная зависимость от случайной переменной $\xi$, т.е. мы включили диагональный замороженный беспорядок типа замещения.
В принципе можно провести аналогичное рассуждение и для случая, когда индекс $i$ нумерует эквивалентные кластеры в решетке и прийти к кластерному обобщению схемы~\cite{Shiba:1971}. Структура получающихся уравнений в этом случае аналогична приведенной выше с заменой скаляров на матрицы там, где это необходимо.

В этих уравнениях $G(z)$~"--- одночастичная электронная функция Грина в одноузельном приближении в спектральном представлении для решеточной задачи в рамках рассматриваемых моделей, $z$ обозначает комплексную энергию, $G^{(i)}[\xi_{i}](z)$~"--- одночастичная электронная функция Грина для узла $i$ в решетке с локальным модельно зависимым взаимодействием $H_{i}[\xi]$ на этом узле и определяемым самосогласованно полем $\Sigma(z)$ на всех остальных узлах решетки, а $G_{0}^{(i)}(z)$~"--- одночастичная электронная функция Грина на узле $i$ системы электронов на решетке взаимодействующих с (одночастичным) полем $\Sigma(z)$, включенном на всех узлах решетки, кроме $i$-того. $V_{i}^{\text{eff}}[\xi_{i}](z)$ является собственно-энергетической частью соответствующей взаимодействию $H_{i}[\xi]$ на узле $i$ в решетке с самосогласованным полем $\Sigma(z)$ на остальных узлах. $V_{i}^{\text{eff}}[\xi_{i}](z)$ определяется только взаимодействием $H_{i}[\xi]$  и для случая локального многоэлектронного взаимодействия и добавленного диагонального замороженного беспорядка типа замещения имеет следующий общий вид
\begin{equation}
V_{i}^{\text{eff}}[\xi_{i}](z)=\xi_{i}+ {\hat{\Sigma}}_{i}[G^{(i)}[\xi_{i}](z)]
\label{eq:5}
\end{equation}
где $\xi_{i}$~"--- случайный потенциал рассеяния на узле $i$, а ${\hat{\Sigma}}_{i}[g]$ некоторый  модельно зависимый локальный одноузельный функционал $g$ соответствующий точному или, на практике, приближенному решению однопримесной задачи без беспорядка и с взаимодействием $H_{i}[0]$ на $i$-том узле. Здесь для простоты мы опустили явную зависимость функционала ${\hat{\Sigma}}_{i}[g]$ от функции Грина подсистемы локальных магнитных моментов для случая $s$"~$f$~обменной модели. Впрочем, учет этого обстоятельства несуществен для наших рассуждений.

Отметим, что если мы выключим многочастичное взаимодействие в системе и оставим только беспорядок, то выписанная система уравнений вместе с уравнением самосогласования представляет собой хорошо известное в теории неупорядоченных систем одноузельное приближение когерентного потенциала (ПКП, CPA).

Для установления связи с решеточными приближениями, о которых мы говорим в следующем разделе, заметим, что для исходной решеточной задачи точное или, что на практике случается чаще, приближенное решение для одночастичной электронной функции Грина определяется уравнением Дайсона с собственно-энергетической частью ${\Sigma}_{ij}(z)={\hat{\Sigma}}_{ij}[G_{kl}]$, где ${\hat{\Sigma}}_{ij}[g_{kl}]$ функционал определяемый только видом взаимодействия в соответствующей решеточной модели. Представим ${\hat{\Sigma}}_{ij}[g_{kl}]$ в виде
\begin{equation}
{\hat{\Sigma}}_{ij}[g_{kl}]={\hat{\Sigma}}_{i}[g_{ii}]\delta_{ij} + \Delta{\hat{\Sigma}}_{ij}[g_{kl}]
\label{eq:6}
\end{equation}
Здесь ${\hat{\Sigma}}_{i}[g_{ii}]$~"---  локальный одноузельный функционал, выделенный из полного нелокального функционала ${\hat{\Sigma}}_{ij}[g_{kl}]$. Таким образом, зная точное или, что важнее, приближенное выражение для ${\hat{\Sigma}}_{ij}[g_{kl}]$ исходной решеточной задачи с локальным многоэлектронным взаимодействием без беспорядка и выделяя соответствующую одноузельную часть, мы можем получить одноузельное приближение для одночастичной электронной функции Грина с учетом беспорядка и тем самым электронный спектр для этого случая.

Для того, чтобы продемонстрировать связь выписанной системы уравненинй (\ref{eq:1})"--~(\ref{eq:5}) с современным подходом DMFT, заметим, что уравнение (\ref{eq:2}) является уравнением Дайсона для эффективного действия системы электронов на одном узле в самосогласованном эффективном поле, определяемом одночастичной электронной функцией Грина $G_{0}^{(i)}(z)$ и локальным многоэлектронным взаимодействием $H_{i}[\xi]$, которое определяет $V_{i}^{\text{eff}}[\xi_{i}](z)$ (см. уравнения (\ref{eq:2}),(\ref{eq:5})). Именно такое действие возникает в DMFT. Тогда, вводя обратную функцию к преобразованию Гильберта
\begin{equation}
\zeta=R[R_{0}(\zeta)] \rightleftharpoons R_{0}(\zeta)=\int\frac{N_{0}(\epsilon)d\epsilon}{\zeta - \epsilon}
\label{eq:7}
\end{equation}
и учитывая (\ref{eq:1}) и (\ref{eq:4}), можно переписать уравнение (\ref{eq:3}) в виде
\begin{equation}
{G_{0}^{(i)}(z)}^{-1}=z + \mu
                        + {{\langle G^{(i)}[\xi](z)\rangle}_{\xi}}^{-1}
                        - R[{\langle G^{(i)}[\xi](z)\rangle}_{\xi}]
\label{eq:8}
\end{equation}
в котором легко узнаем уравнение самосогласования DMFT в том виде, в котором оно возникает в этой теории (мы также явно выписали зависимость от химпотенциала). Таким образом, исходная система уравнений одноузельного приближения (\ref{eq:1})"--~(\ref{eq:5}) эквивалентна системе (\ref{eq:2}), (\ref{eq:5}), (\ref{eq:7}) и (\ref{eq:8}). Последняя система уравнений и является системой самосогласованных уравнений DMFT; именно в таком виде она и возникает в этой теории. Это позволяет охарактеризовать метод DMFT как обобщение приближения когерентного потенциала на системы с многоэлектронным взаимодействием или как многочастичное ПКП;  в  отсутствие многоэлектронного взаимодействия система уравнений (\ref{eq:2}), (\ref{eq:5}), (\ref{eq:7}) и (\ref{eq:8}) представляет собой формулировку ПКП в форме приближения DMFT. Решение же системы (\ref{eq:1})"--~(\ref{eq:5}) с некоторым приближением для $V_{i}^{\text{eff}}[\xi_{i}](z)$ есть одновременно решение эквивалентной задачи DMFT, причем в том же самом приближении.

Из приведенного обсуждения также ясно, что решеточные приближения для полного функционала ${\hat{\Sigma}}_{ij}[g_{kl}]$ можно использовать после выделения одноузельных вкладов в качестве сольверов эффективной нульмерной задачи DMFT.

Систему уравнений (\ref{eq:1})"--~(\ref{eq:5}) можно представить в виде, обычном для ПКП~"--- через одноузельную $T$-матрицу. Для этого с помощью уравнения (\ref{eq:3}) выразим $G_{0}^{(i)}(z)$ и подставим в уравнение (\ref{eq:2}). Тогда получим

\begin{equation}
G^{(i)}[\xi_{i}](z)=G(z) + G(z)\left(V_{i}^{\text{eff}}[\xi_{i}](z)-\Sigma(z)\right)G^{(i)}[\xi_{i}](z)
\label{eq:9}
\end{equation}

Введем далее стандартным соотношением одноузельную $T$-матрицу
\begin{equation}
 t(z)G(z) = \left(V_{i}^{\text{eff}}[\xi_{i}](z)-\Sigma(z)\right)G^{(i)}[\xi_{i}](z)
\label{eq:10}
\end{equation}
тогда для $t(z)$ имеем явное выражение
\begin{equation}
t(z)=\frac{
         \left(V_{i}^{\text{eff}}[\xi_{i}](z)-\Sigma(z)\right)
          }
          {
         1 - \left(V_{i}^{\text{eff}}[\xi_{i}](z)-\Sigma(z)\right)G(z)
          }
\label{eq:11}
\end{equation}
и уравнение (\ref{eq:9}) запишется в виде
\begin{equation}
G^{(i)}[\xi_{i}](z)=G(z) + G(z)t(z)G(z)
\label{eq:12}
\end{equation}
Усредняя по беспорядку и используя (\ref{eq:4}), получаем условие самосогласования через $T$-матрицу
\begin{equation}
{\langle t(z)\rangle}_{\xi}=0
\label{eq:13}
\end{equation}
В  отсутствие беспорядка имеем очевидное решение $\Sigma(z)=V^{\text{eff}}_{i}[G(z)]$. Таким образом, если $V^{\text{eff}}_{i}[G(z)]$~"--- функционал ужирненной локальной функции Грина, то решение исходной решеточной модели, решение системы для одноузельного приближения и решение эффективной задачи DMFT полностью совпадают. Поэтому проведенное рассмотрение может быть полезно для обсуждения как идеальных систем, так и систем с беспорядком. Более того, для такого обсуждения систем с беспорядком, как ясно из построения одноузельного приближения, можно использовать локальное самосогласованное приближение для решеточной задачи без беспорядка.

\section{Самосогласованные уравнения для функции Грина}

Несамосогласованные выражения в рассматриваемых приближениях функции Грина в модели Хаббарда и $s$"~$d$~обменной модели обладают рядом недостатков (в частности, не позволяют корректно описать переход металл---изолятор).

В работе \cite{Anokhin:1991:OTM} были рассмотрены и проанализированы  ряд самаосогласованных приближений в модели Хаббарда и квазиклассической $s$"~$d$~обменной модели типа ПКП (в частности, расцепление "<Хаббард-III">), а также более простые.

Для~получения самосогласованного приближения в~выражениях для~функции Грина (\ref{eq:HM:GF:1}) в модели Хаббарда заменим локаторы в~приближении "<Хаббард"~I"> на~самосогласованные (ср.~\cite{Irkhin:2001:MIT}), т.~е.
\begin{equation}
G_{\mathbf{q}}^{0}(E)\rightarrow G_{\mathbf{q}}(E).
\label{eq:HM:s-c}
\end{equation}

Для $s$"~$d$~обменной модели применялось три разных схемы получения самосогласованных уравнений для функций Грина.

Приближение~(\ref{eq:sd:a0:loc}), (\ref{eq:sd:b0:loc}) для $s$"~$d$~обменной модели обладает неприятным дефектом~"---  плотность состояний имеет особенности типа ван~Хова при энергиях, соответствующих краям зоны в приближении "<Хаббард"~I">~\cite{Anokhin:1991:ICC,Irkhin:1999:KEP}. Чтобы избавиться от этого недостатка, мы
можем самосогласованно перенормировать ширину зоны
\begin{equation}
P_{\sigma \alpha }\rightarrow \widetilde{P}_{\sigma \alpha }
\label{eq:sd:s-c:1}
\end{equation}
в выражениях для функций Грина~(\ref{eq:sd:EGF:H1P}) и функций распределения, входящих в выражения~(\ref{eq:sd:a0}) и~(\ref{eq:sd:b0}) (первая схема самосогласования). Благодаря этой процедуре, края зоны в резольвентах совпадают с краями для функции Грина~(\ref{eq:sd:EGF:0}). Это приближение сохраняет квазичастичную картину, в отличие от приближения "<Хаббард"~III"> и самосогласованных приближений, рассматриваемых ниже, где получается сильно некогерентное поведение.

Вторая схема самосогласования заключается в замене в знаменателей дробей подинтегральных выражений~(\ref{eq:sd:a0}) и~(\ref{eq:sd:b0}) по правилу~\cite{Irkhin:1999:KEP}
\begin{equation}
E-P_{\sigma \alpha } t_{\mathbf{q}}\rightarrow b_{\mathbf{q}\sigma \alpha }(E)-a_{\mathbf{q}\sigma \alpha }(E)t_{\mathbf{q}}.
\label{eq:sd:s-c:2}
\end{equation}

В третьей, более последовательной схеме построения самосогласованного приближения функции Грина~(\ref{eq:sd:EGF:0}), входящие в выражения~(\ref{eq:sd:a0:loc}) и~(\ref{eq:sd:b0:loc}) и соответствующие приближению "<Хаббард"~I">, заменяются самосогласованными величинами~(\ref{eq:sd:EGF:0}), т.~е.
\begin{equation}
G_{\mathbf{q}\sigma \alpha }^{0}(E)\rightarrow G_{\mathbf{q}\sigma \alpha }(E),
\label{eq:sd:s-c}
\end{equation}
и совпадает со схемой самосогласования~(\ref{eq:HM:s-c}), используемой нами для модели Хаббарда.

Здесь мы, чтобы получить локальное самосогласованное приближение в полученных промежуточных выражениях~(\ref{eq:HM:A:1}) и~(\ref{eq:HM:B:1})  проведем замену несамосогласованного локатора на точный по схеме~(\ref{eq:HM:s-c}) и пренебрежем импульсной зависимостью спиновых корреляторов.
Тогда самосогласованная функция Грина в модели Хаббарда при конечных $U$ и половинном заполнении зоны запишется как (ср.(\ref{eq:HM:A:1})-(\ref{eq:HM:B:1}))
\begin{equation}
G_{\mathbf{k}}(E)=[F(E)-t_{\mathbf{k}}]^{-1},\quad
F(E)=\frac{B(E)}{A(E)},
\label{eq:HMHF:GF:1:s-c}
\end{equation}
где
\begin{eqnarray}
A(E) &=&1+\frac{U^{2}}{E^{2}}
\sum_{\mathbf{q}}\chi t_{\mathbf{q}}G_{\mathbf{q}}(E)
+\frac{2U}{E}\sum_{\mathbf{q}}t_{\mathbf{q}}n_{\mathbf{q}}G_{\mathbf{q}}(E),
\label{eq:HMHF:A:1} \\
B(E) &=&F^{0}(E)+\frac{2U}{E}\sum_{\mathbf{q}}t_{\mathbf{q}}^{2}
n_{\mathbf{q}}G_{\mathbf{q}}(E).
\label{eq:HMHF:B:1}
\end{eqnarray}

Для $s$"~$d$~обменной модели в случае конечных~$I$ применялась вторая схема самосогласования~(\ref{eq:sd:s-c:2}), предложенная в \cite{Anokhin:1991:OTM}.
Этот вариант самосогласования приводит к следующему выражению для величины $A_{\mathbf{k}}(E)$~(см. (\ref{eq:a:sd})):
\begin{equation}
 A(E)=1+{\left(\frac{IS}{E}\right)}^2
         \sum_{\mathbf{q}} \frac{\chi}{(2S)^2}
         \frac{t_\mathbf{q}}{y(E)-A(E)t_{\mathbf{q}}}.
\label{eq:sdii:A:1:s-c}
\end{equation}
Здесь, как и в случае для модели Хаббарда, для получения локального приближения мы пренебрегли зависимостью
магнитной восприимчивости от квазиимпульса.

Применение схемы самосогласования типа "<Хаббард"~III"> в квазиклассической $s$"~$d$~обменной модели дает результат
\begin{equation}
F(E)=\frac{y(E)}{a(E)}=\frac{y(E)-\lambda (E)}{1-\lambda (E)/E},
\end{equation}
\begin{equation}
\lambda (E)=F(E)-{\left( \frac{\chi }{S^{2}} \sum_{\mathbf{q}} G_{\mathbf{q}}(E) \right)}^{-1}
\end{equation}


Для $s$"~$d$~обменной модели с $|I|=\infty $ при численных расчетах плотности состояний применялись все три схемы самосогласования~(\ref{eq:sd:s-c:1}),(\ref{eq:sd:s-c:2}) и~(\ref{eq:sd:s-c}), однако здесь мы приведем формулы только для третьей, более последовательной. Тогда выражения (\ref{eq:sd:a0:loc})---(\ref{eq:sd:b0:loc}) в пренебрежении квазиимпульсной зависимостью спиновых корреляторов приобретают  вид
\begin{eqnarray}
A_{\sigma \alpha }(E) &=&P_{\sigma \alpha }+\sum_{\mathbf{q}}\frac{t_{\mathbf{k-q}}}{(2S+1)^{2}}\frac{\chi ^{\sigma -\sigma }+(2S+1)n_{\mathbf{k-q}-\sigma \alpha }}{P_{-\sigma \alpha }}G_{\mathbf{k-q}-\sigma \alpha }(E)+
 \nonumber \\
&&+\sum_{\mathbf{q}}\frac{t_{\mathbf{k-q}}}{(2S+1)^{2}}\frac{\chi ^{zz}}{P_{\sigma \alpha }}G_{\mathbf{k-q}\sigma \alpha }(E),
 \label{eq:sd:A:loc} \\
B_{\sigma \alpha }(E) &=&E-\sum_{\mathbf{q}}\frac{t_{\mathbf{k-q}}^{2}}{2S+1}\frac{n_{\mathbf{k-q}-\sigma \alpha }}{P_{-\sigma \alpha }}G_{\mathbf{k-q}-\sigma \alpha },
 \label{eq:sd:B:loc}
\end{eqnarray}
Мы не приводим явных выражений для этих функций в случае модели Хаббарда с $U=\infty $, поскольку они могут быть получены из эквивалентности модели Хаббарда с $U=\infty $ и~$s$"~$d$~обменной модели с $I=-\infty $ и~$S=1/2$ с помощью обсуждавшихся ранее замен~(\ref{eq:HM2sd}).

Общей особенностью рассмотренных приближений является их (квази)локальность в том смысле, что все они имеют общую структуру обратного локатора вида $F(E)=\hat{F}[G_{\text{loc}}(E)](E)$, т.~е. обратный локатор функционально зависит только от локальной функции Грина.

\section{Результаты и обсуждение}

Представленные рассуждения продемонстрированы на полуэллиптической затравочной плотности состояний (решетка Бете с бесконечным координационным числом). Расчеты также проводились и для прочих модельных и реалистичных функций плотностей состояний~\cite{Irkhin:1999:KEP,Irkhin:2001:MIT} с особенностями ван~Хова (последние приводят к дополнительному усложнению картины).

\begin{figure}[p]
\begin{center}
\includegraphics[width=0.45\textwidth]{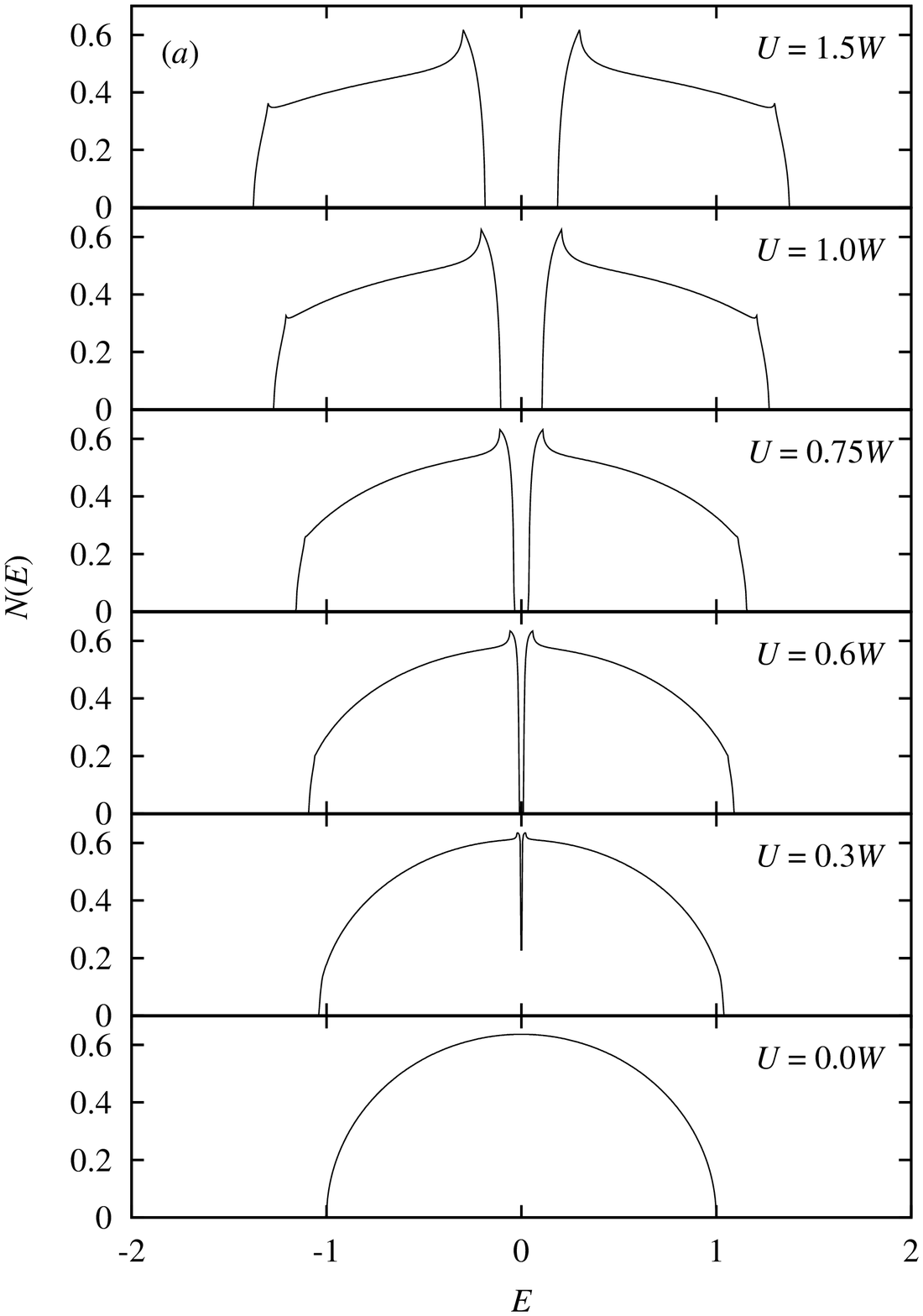}
\includegraphics[width=0.45\textwidth]{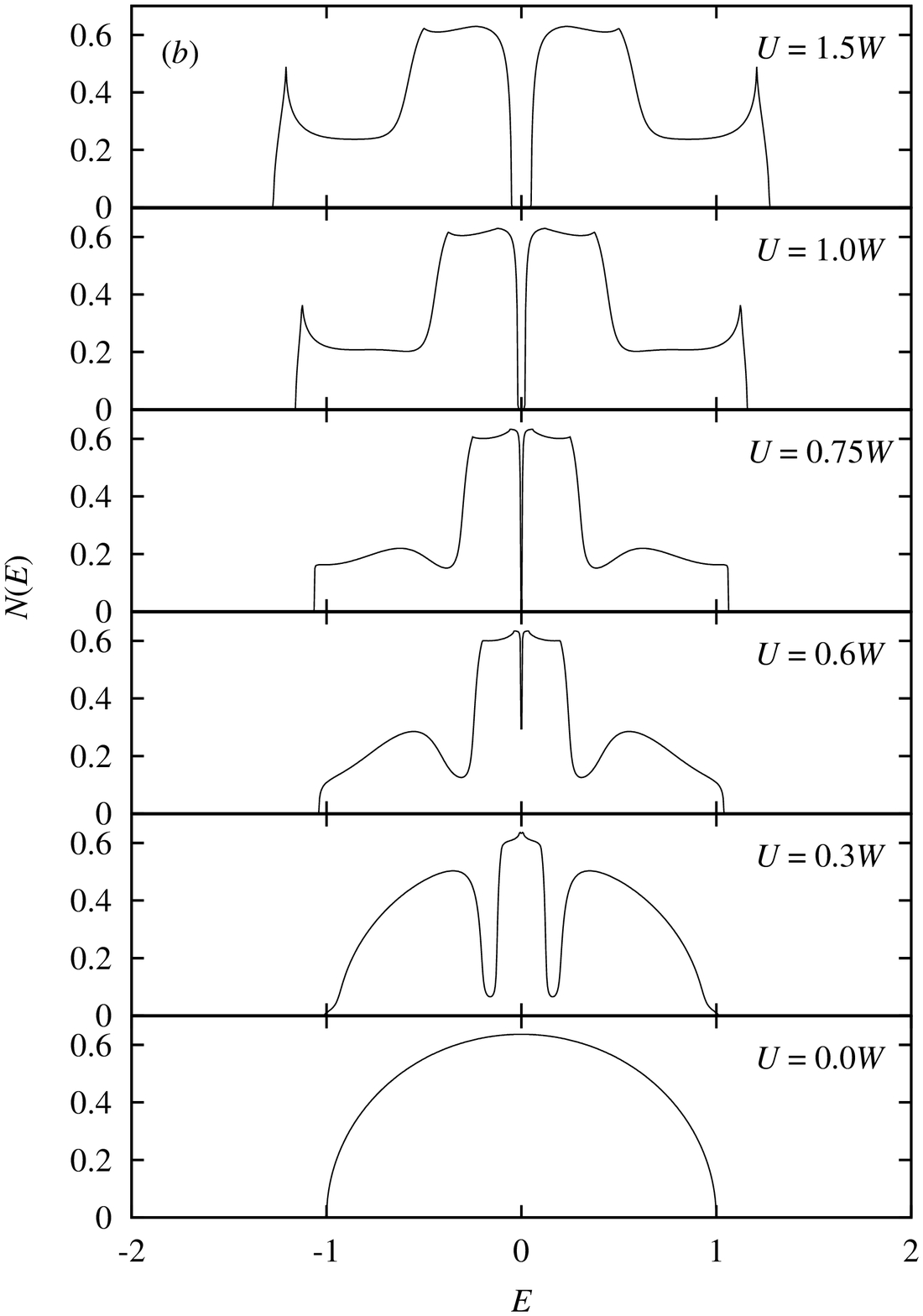}
\end{center}
\caption{Плотность состояний для полуэллиптической затравочной зоны при различной величине хаббардовского параметра в несамосогласованном приближении. Рисунок (\textit{a}) соответствуют приближению~(\ref{eq:b:sd})"--~(\ref{eq:a:sd}), (\textit{b})~"--- (\ref{eq:HM:A:1})---(\ref{eq:HM:B:1})}
\label{fig:01}
\end{figure}

\begin{figure}[p]
\begin{center}
\includegraphics[width=0.45\textwidth]{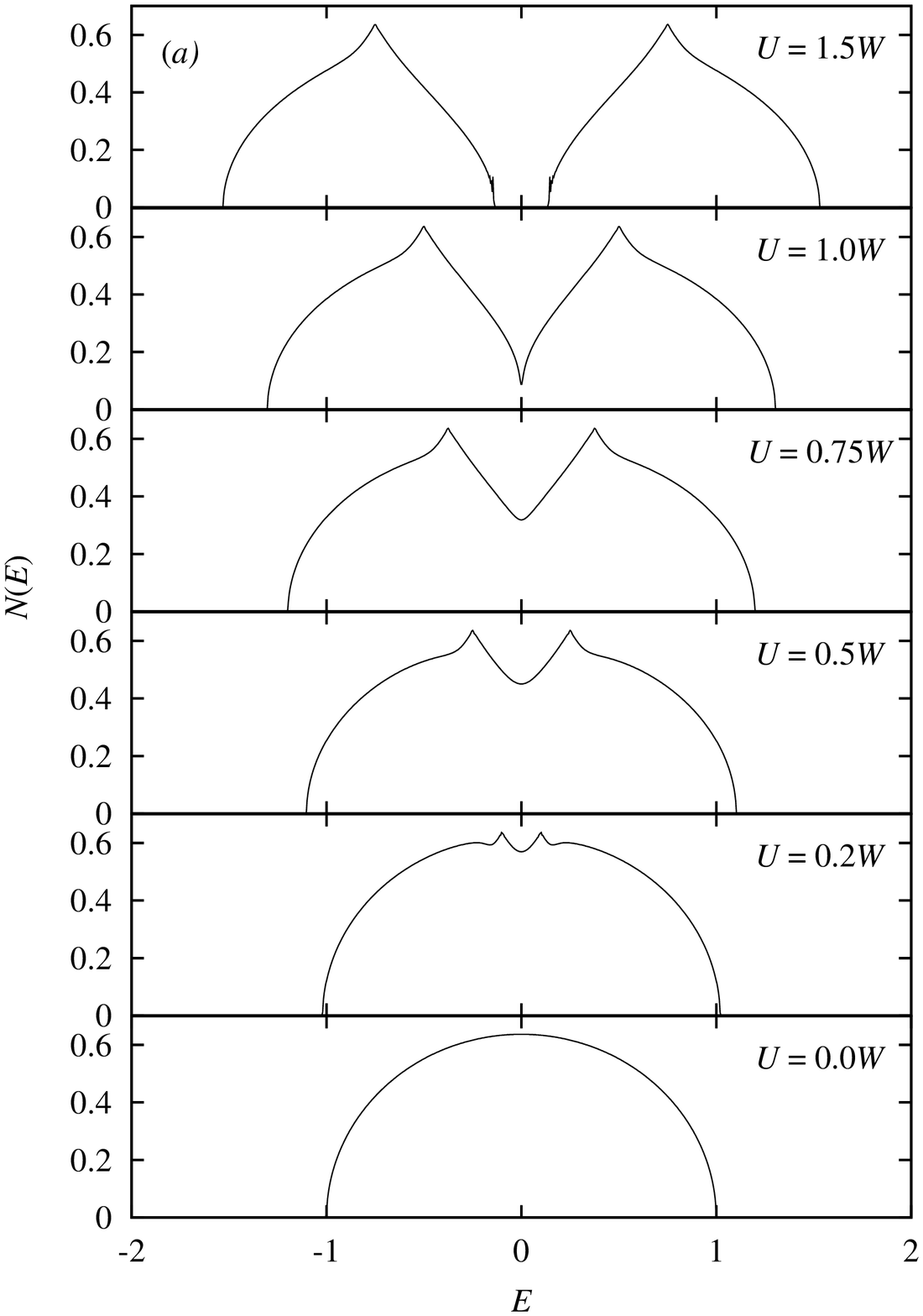}
\includegraphics[width=0.45\textwidth]{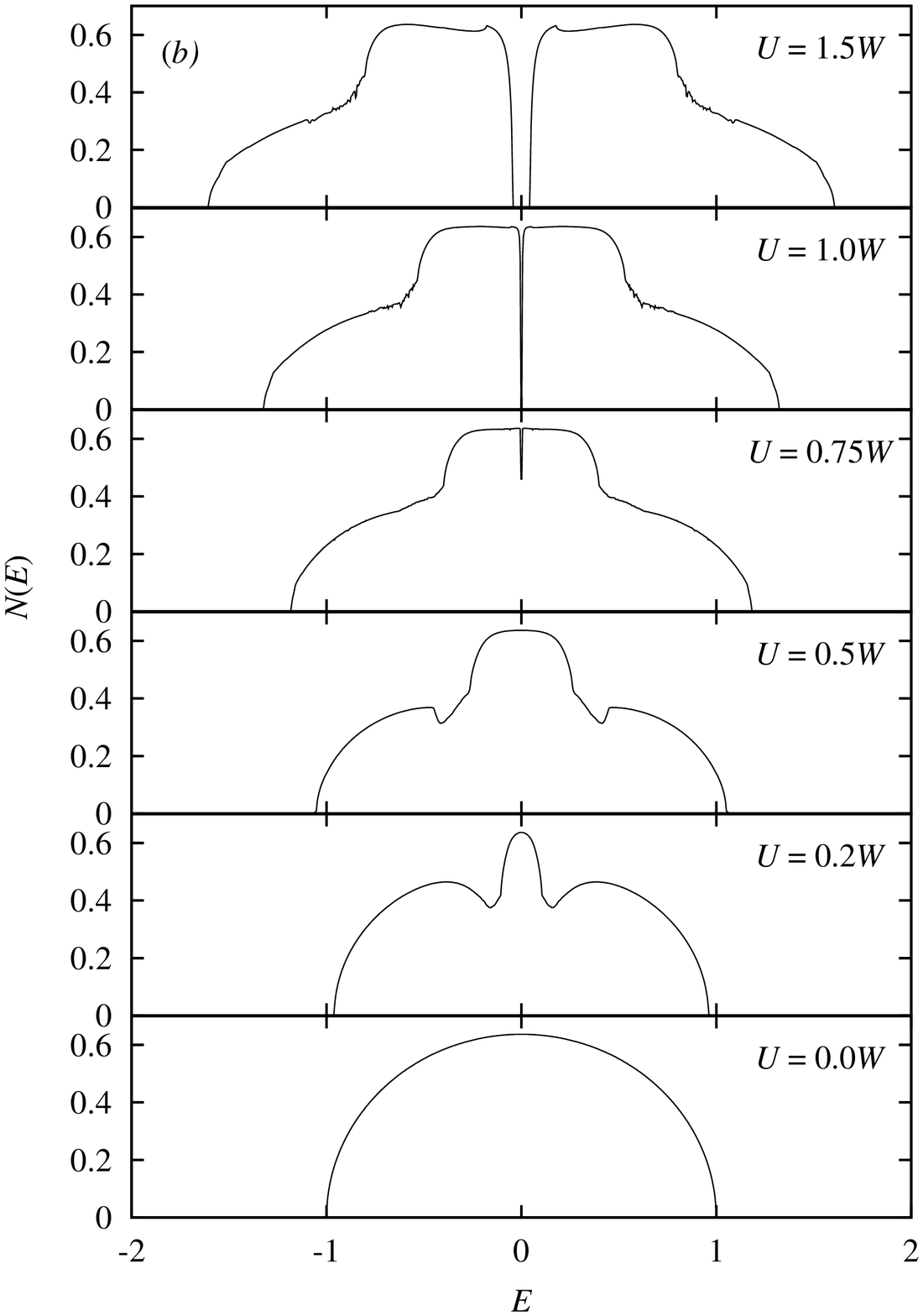}
\end{center}
\caption{Плотность состояний полуэллиптической затравочной зоны при различной величине хаббардовского параметра в самосогласованном приближении. Рисунок (\textit{a}) соответствуют приближению~(\ref{eq:sdii:A:1:s-c}), (\textit{b})~"--- (\ref{eq:HMHF:A:1})---(\ref{eq:HMHF:B:1})}
\label{fig:02}
\end{figure}

\begin{table}[p]
\caption{Критические величины кулоновского отталкивания перехода металл"--~изолятор для различных затравочных плотностей состояний в~приближении "<Хаббард-III"> ($U_{{\rm c}}^{{\rm H}}$)~\cite{Hubbard-III:1964,Anokhin:1991:OTM}, из~"<линеаризованной"> динамической теории среднего поля~($U_{{\rm c}}^{{\rm L}}$)~\cite{Bulla:1999:LDMFT}, а~также в~нашем несамосогласованном~($U_{{\rm c}}^{{\rm NSC}}$) и~самосогласованном ($U_{{\rm c}}^{{\rm SC}}$) приближениях}
\label{tab:3.1}
\begin{center}
\small
\begin{tabular}{|l|cc|cc|}
\hline
             & $U_{{\rm c}}^{{\rm H}}/W$ & $U_{{\rm c}}^{{\rm L}}/W$ & $U_{{\rm c}}^{{\rm NSC}}/W$ & $U_{{\rm c}}^{{\rm SC}}/W$ \\ \hline
прямоугольная     & $1$                       & $1{,}73$                  & $0{,}99$                    & $1{,}22$                   \\
полуэллиптическая & $\sqrt{3}/2=0{,}866$      & $1{,}5$                   & $0{,}87$                    & $1{,}06$                   \\
гауссова          & $\sqrt{3}/2=0{,}866$      & $1{,}5$                   & $0{,}87$                    & $1{,}06$                   \\
квадратная        & $0{,}866$                 & $1{,}5$                   & $0{,}87$                    & $1{,}06$                   \\
п.~к.             & $0{,}707$                 & $1{,}22$                  & $0{,}76$                    & $0{,}99$                   \\
о.~ц.~к.          & $0{,}612$                 & $1{,}06$                  & $0{,}67$                    & $0{,}92$                   \\ \hline
\end{tabular}
\normalsize
\end{center}
\end{table}

\subsection{Переход металл"--~изолятор}

В модели Хаббарда с половинным заполнением зоны имеет место переход металл--изолятор при $U$ порядка ширины зоны~$W$. Соответствующие критические величины кулоновского отталкивания приведены в таблице \ref{tab:3.1}.

Критические величины в~самосогласованном приближении~(\ref{eq:HMHF:GF:1:s-c}) несколько изменяются по~сравнению с~результатами в~приближении "<Хаббард-III">, что отражено в~табл.~\ref{tab:3.1}. В~отличие от~результатов работы~\cite{Luo:2000}, где~критическая величина уменьшается при~учёте флуктуаций~(для решётки Бете $U_{\mathrm{c}}/W=0{,}67)$, рассматриваемый здесь подход даёт противоположный эффект, что согласуется с~результатами вычислений квантовым методом Монте-Карло при~конечных температурах, $U_{\mathrm{c}}/W\approx 1$~(см.~табл.~\ref{tab:3.2}). Аналитическое выражение для~$U_{\mathrm{c}}$ в~"<линеаризованной"> теории динамического среднего поля~\cite{Bulla:1999:LDMFT,Ono:2000,Ono:2001} даёт б\'{о}льшие значения, чем в~приближении "<Хаббард-III">,
\begin{equation}
U_{\mathrm{c}}^{\mathrm{L}}
=\sqrt{3}U_{\mathrm{c}}^{\mathrm{H}}=6\sqrt{\mu _{2}},
\label{eq:HMHF:Uc:LDMFT}
\end{equation}
где $\mu _{2}$~"--- второй момент затравочной плотности состояний, ср.~\cite{Irkhin:2001:MIT}.

Сравнивая результаты, приведённые в~табл.~\ref{tab:3.2}, можно
сказать, что это приближение несколько завышает критическую
величину~$U_{\mathrm{c}}$.

\begin{table}[p]
\caption{Критические величины кулоновского отталкивания перехода металл"--~изолятор для решётки Бете~($U_{{\rm c}}^{{\rm B}}$) и~гиперкубической решётки с большими~$d$~($U_{{\rm c}}^{{\rm G}}$) из~работ различных авторов. Некоторые подходы дают два критических значения $U_{\rm c1,\,2}$, соответствующих появлению затухания на поверхности Ферми (плохой металл) и щели (переход в диэлектрическую фазу)}
\label{tab:3.2}
\begin{center}
\small
\begin{tabular}{|llll|ll|}
\hline
$U_{{\rm c1}}^{{\rm B}}/W$ & $U_{{\rm c2}}^{{\rm B}}/W$ & $U_{{\rm c1}}^{{\rm G}}/W$ & $U_{{\rm c2}}^{{\rm G}}/W$ & Ссылка                                & Метод \\ \hline
                           &                            & $1{,}202$                  &                            & \cite{Jarrell:1992,Jarrell:1992:MDP} & QMC   \\
                           &                            & $1{,}24$                   &                            & \cite{Pruschke:1993}                  & QMC   \\
                           & $1{,}685$                  &                            &                            & \cite{Zhang:1993}                     & PT, QMC \\
$1{,}308$                  & $1{,}591$                  & $1{,}273$                  & $1{,}662$                  & \cite{Georges:1993}                   & MFT, IPT \\
                           &                            & $1{,}273$                  &                            & \cite{Caffarel:1994}                  & QMC \\
$1{,}3$                    & $1{,}64$                   &                            &                            & \cite{Rozenberg:1994}                 & MFT, QMC \\
                           & $1{,}45$                   &                            &                            & \cite{Moeller:1995}                   & PSCA \\
$1{,}25$                   &                            &                            &                            & \cite{Schlipf:1999}                   & QMC \\
$1{,}25$                   & $1{,}47$                   & $1{,}15$                   & $1{,}45$                   & \cite{Bulla:1999,Bulla:2000}         & DMFT, NRG, IPT \\
$1{,}195$                  & $1{,}47$                   &                            &                            & \cite{Bulla:2000:FTN}                 & NRG \\
$0{,}67$                   &                            &                            &                            & \cite{Luo:2000}                       & IHIII \\ \hline
\end{tabular}
\normalsize
\end{center}
\textit{QMC~"--- Quantum Monte Carlo Method, квантовый метод Монте-Карло,
PT~"--- Perturbation Theory, теория возмущений,
IPT~"--- Iterative Perturbation Theory, итеративная теория возмущений,
MFT~"--- Mean Field Theory, теория среднего поля,
DMFT~"--- Dynamical Mean Field Theory, динамическая теория среднего поля,
PSCA~"--- Projective Self-Consistent Approximation, проективное самосогласованное приближение,
NRG~"--- Numerical Renormalization Group Method, метод численной  ренормгруппы,
IHIII~"--- Improved Hubbard~III Approximation, улучшенное приближение "<Хаббард-III">.}
\end{table}

Учёт фермиевских возбуждений (который не был проведен в работе~\cite{Anokhin:1991:OTM}) приводит к~изменению вида плотности состояний, в частности к возникновению пика в центе зоны (рис.~\ref{fig:02}). По~сравнению с~приближением "<Хаббард-III"> в~рассматриваемом случае заметно существование псевдощели вблизи перехода металл"--~изолятор при~$U<U_{\mathrm{c}}$. Та~же самая особенность может~быть отмечена в~результатах работы~\cite{Luo:2000}. Из рис.~\ref{fig:01} видно, что при учете фермиевских возбуждений переход металл"--~изолятор возникает даже в несамосогласованном приближении, хотя в этом случае его описание видимо некорректно.

Трёхпиковая структура плотности состояний~(см.~рис.~\ref{fig:01}) имеется при малых~$U$ и размывается при~приближении к~переходу металл"--~изолятор (центральный пик становится широким и~появляется псевдощель). Детали картины перехода металл"--~изолятор в~нашем случае отличны от~картины "<линеаризованной"> теории динамического среднего поля, где~центральный квазичастичный пик постепенно сужается и~исчезает при~$U\rightarrow U_{\mathrm{c}}-0$. Вероятно, это различие связано с~переоценкой роли затухания в~нашем методе. Последовательная трактовка затухания является трудной проблемой. Чтобы обойти эту трудность, большинство вычислений в~случае больших~$d$ выполняется при~конечных температурах. Важно отметить, что различные версии вычислений в~рамках "<линеаризованной"> теории динамического среднего поля дают отчасти разные картины перехода металл"--~изолятор. Отметим, что расчеты методом Монте-Карло не демонстрируют тенденцию формирования пика в центре зоны (см., напр.,~\cite{Laad:2001}).


\subsection{$s-d$ обменная модель с $|I|=\infty $}

\begin{figure}[p]
\begin{center}
\includegraphics[width=0.45\textwidth]{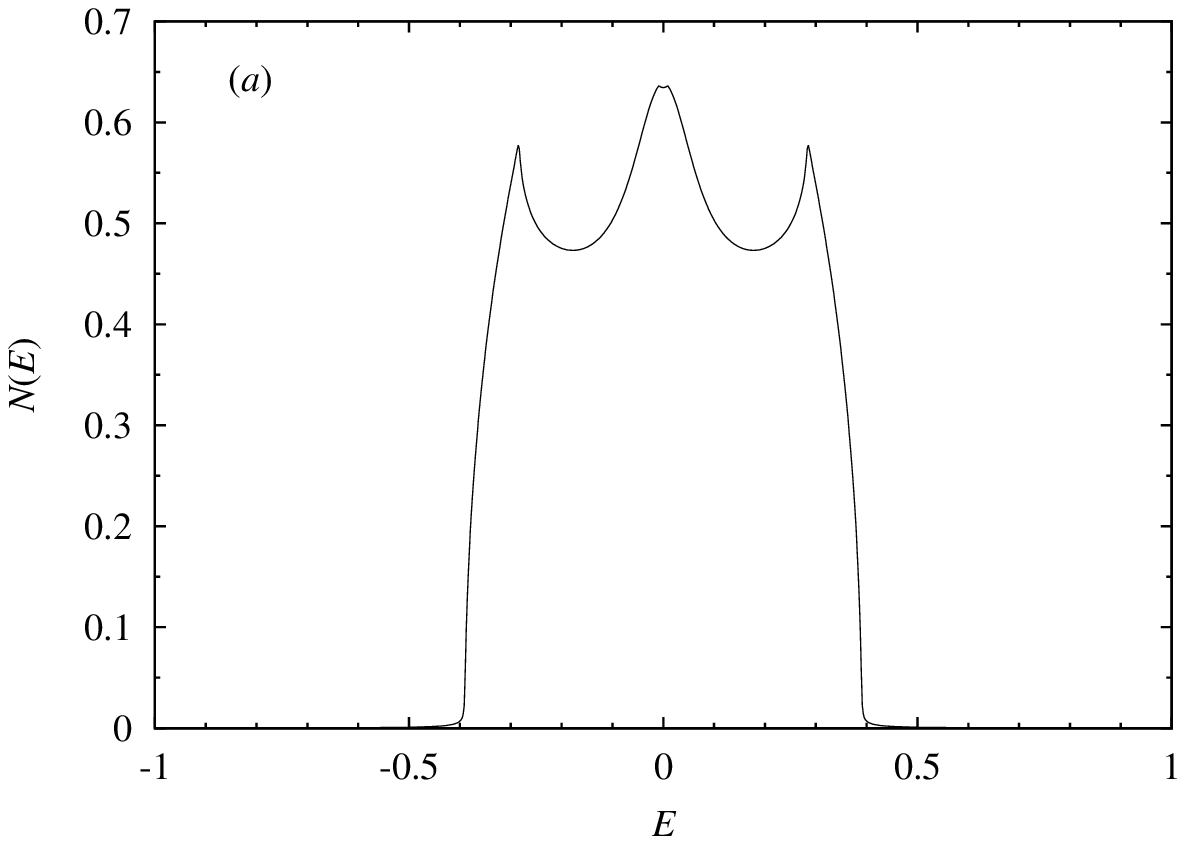}
\includegraphics[width=0.45\textwidth]{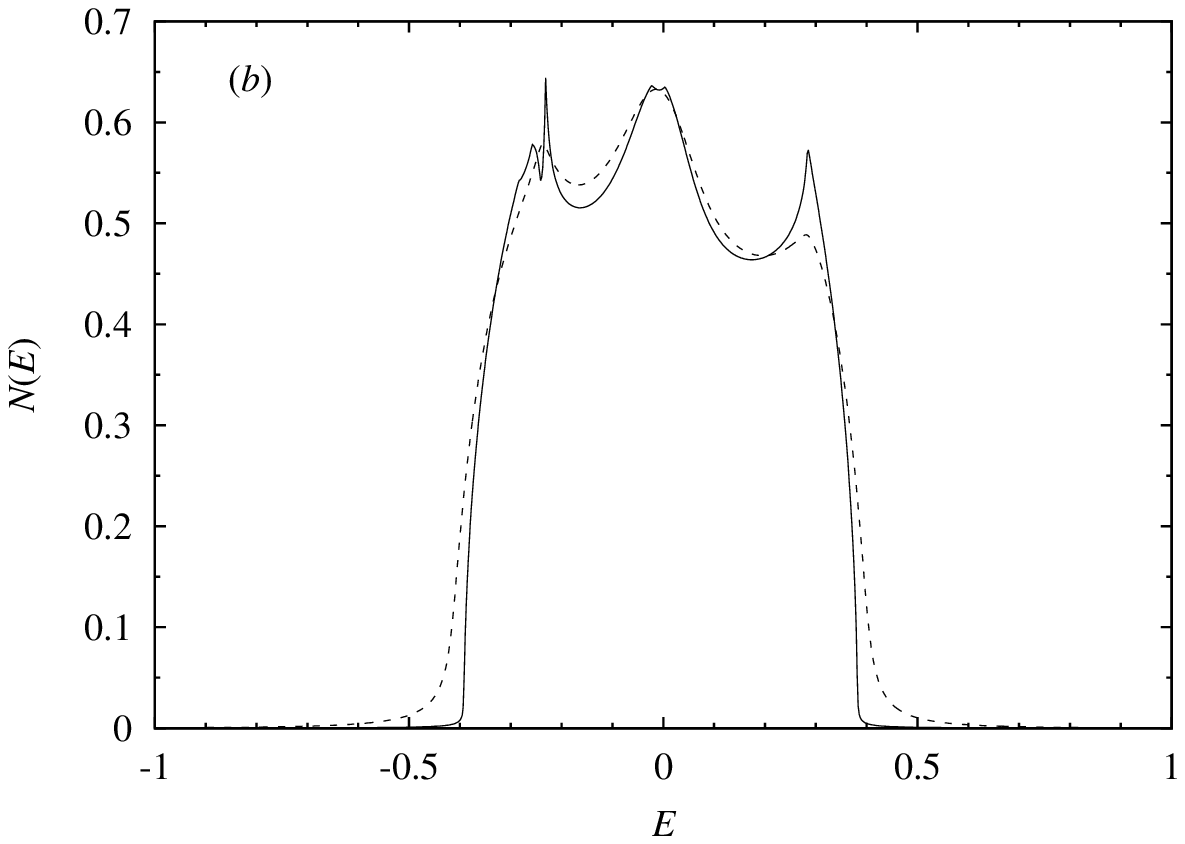}
\includegraphics[width=0.45\textwidth]{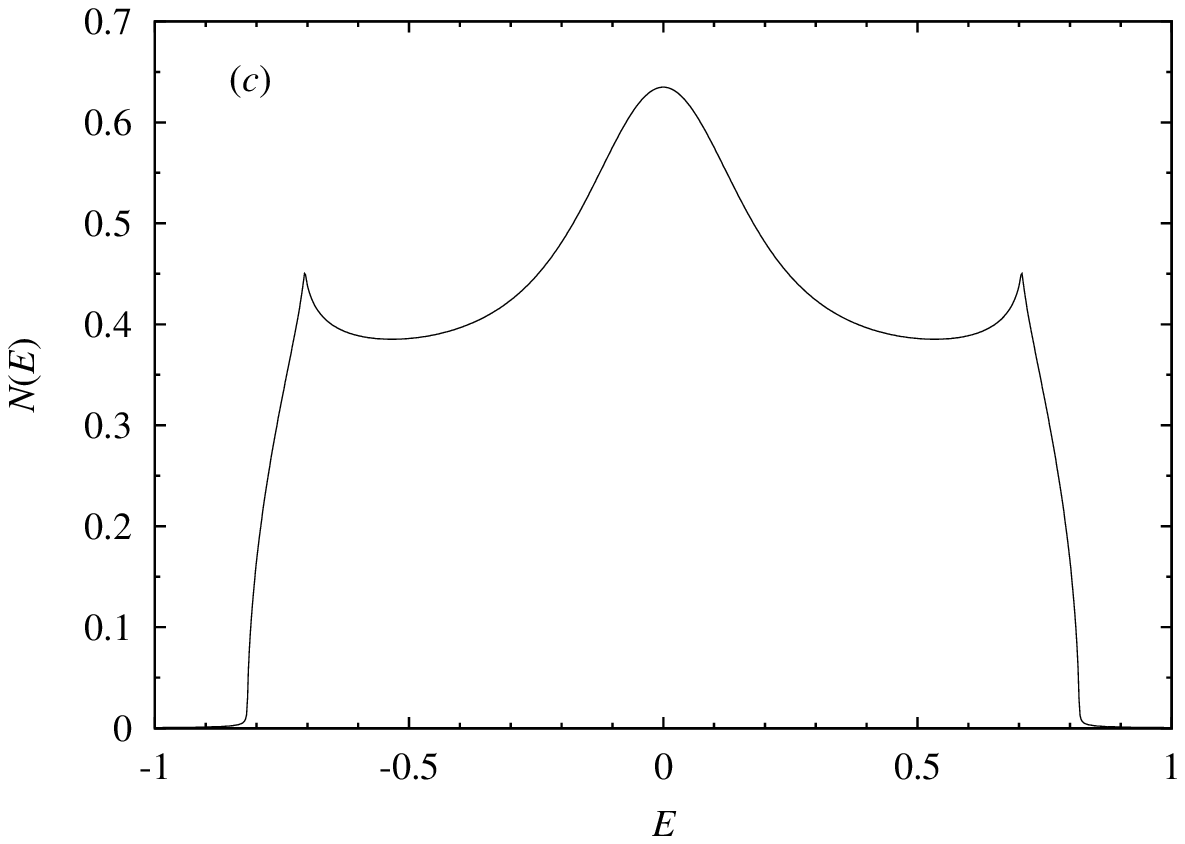}
\includegraphics[width=0.45\textwidth]{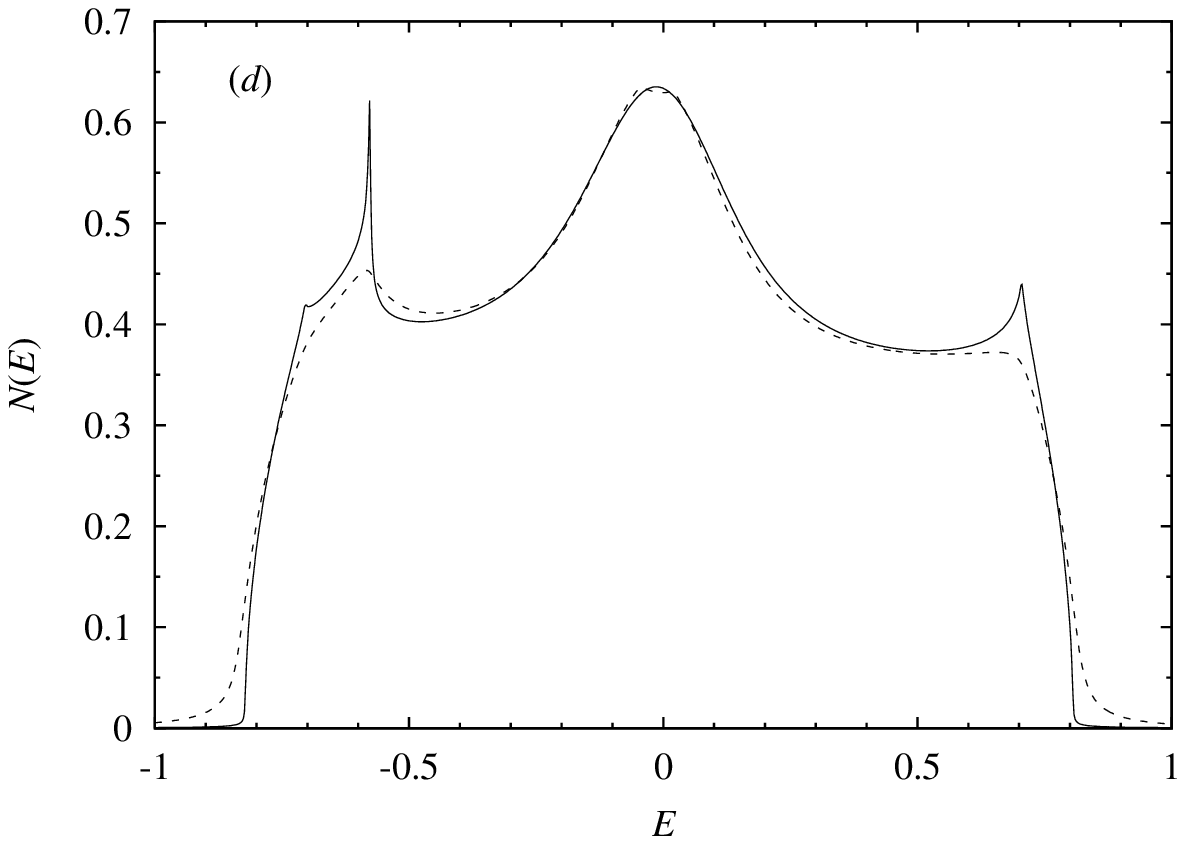}
\end{center}
\caption{Плотность состояний для затравочной полуэллиптической зоны  для $s$"~$d$~обменной модели при~$S=1/2$ $n=0$ и~$n=0{,}15$ для~$\alpha =+$ (\textit{a}, \textit{b}) и~$\alpha =-$~(\textit{c}, \textit{d}) в несамосогласованном приближении~(\ref{eq:sd:EGF:0}). Сплошная линия соответствует температуре~$T=0$, без учёта спиновой динамики~(\ref{eq:sd:a0:loc})---(\ref{eq:sd:b0:loc}), а штриховая линия~"--- при нулевой температуре, с учётом спиновой динамики~($\omega _{\max }=0{,}257n$)~(\ref{eq:sd:A:K})---(\ref{eq:sd:B:K}). Энергия и температура измеряются в единицах полуширины затравочной зоны}
\label{fig:03}
\end{figure}

\begin{figure}[p]
\begin{center}
\includegraphics[width=0.45\textwidth]{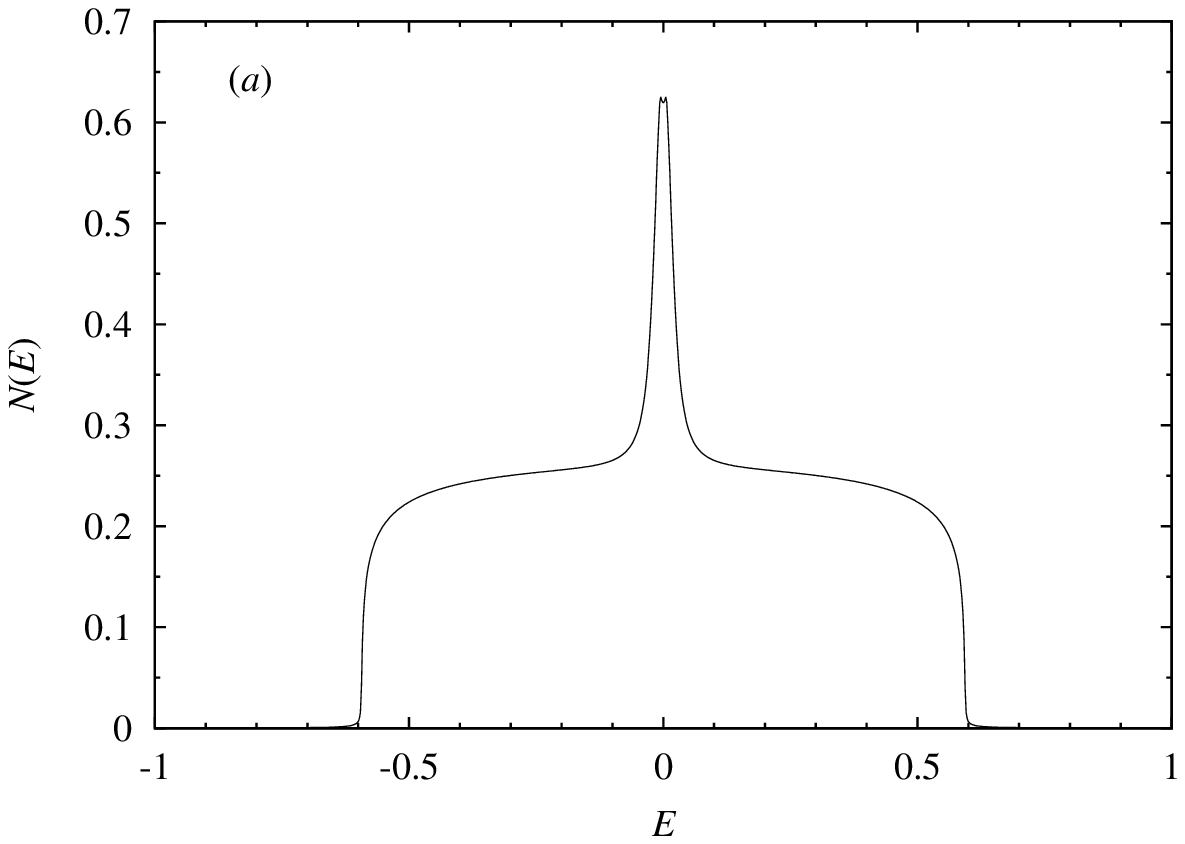}
\includegraphics[width=0.45\textwidth]{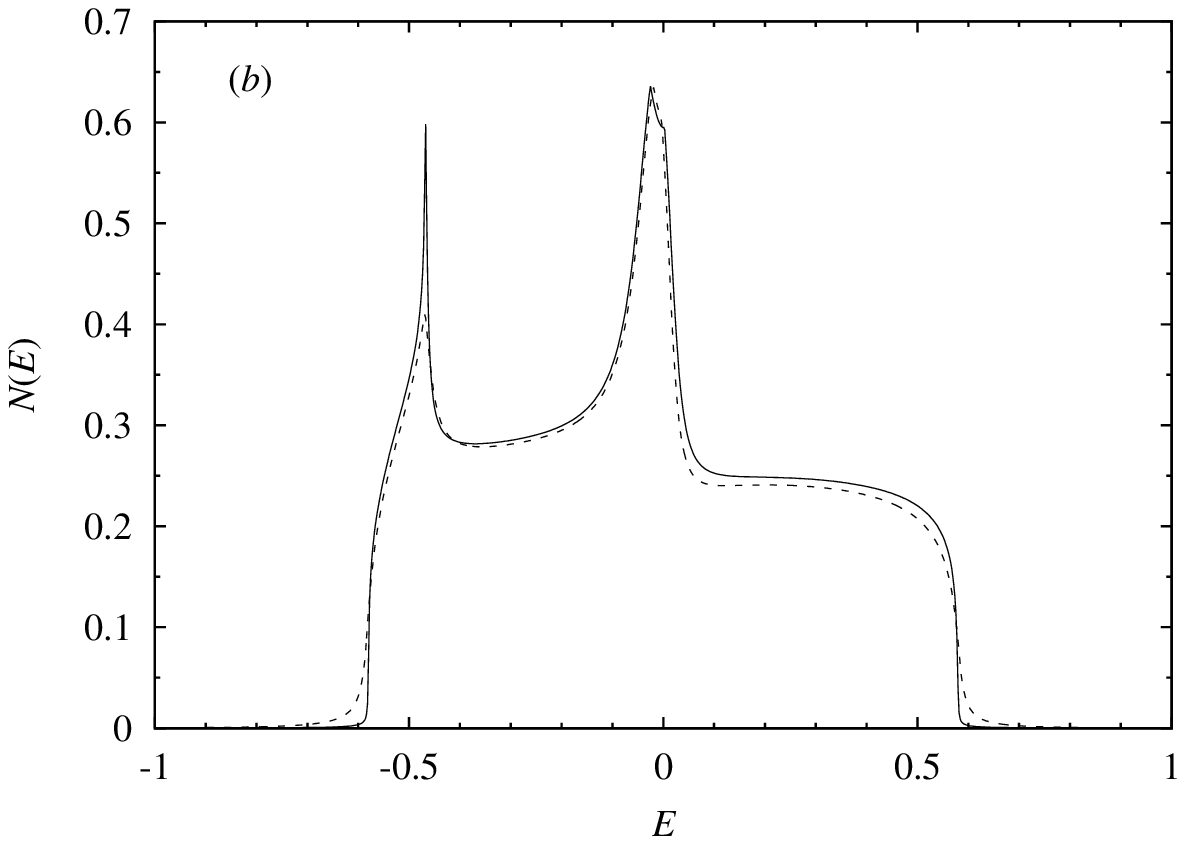}
\includegraphics[width=0.45\textwidth]{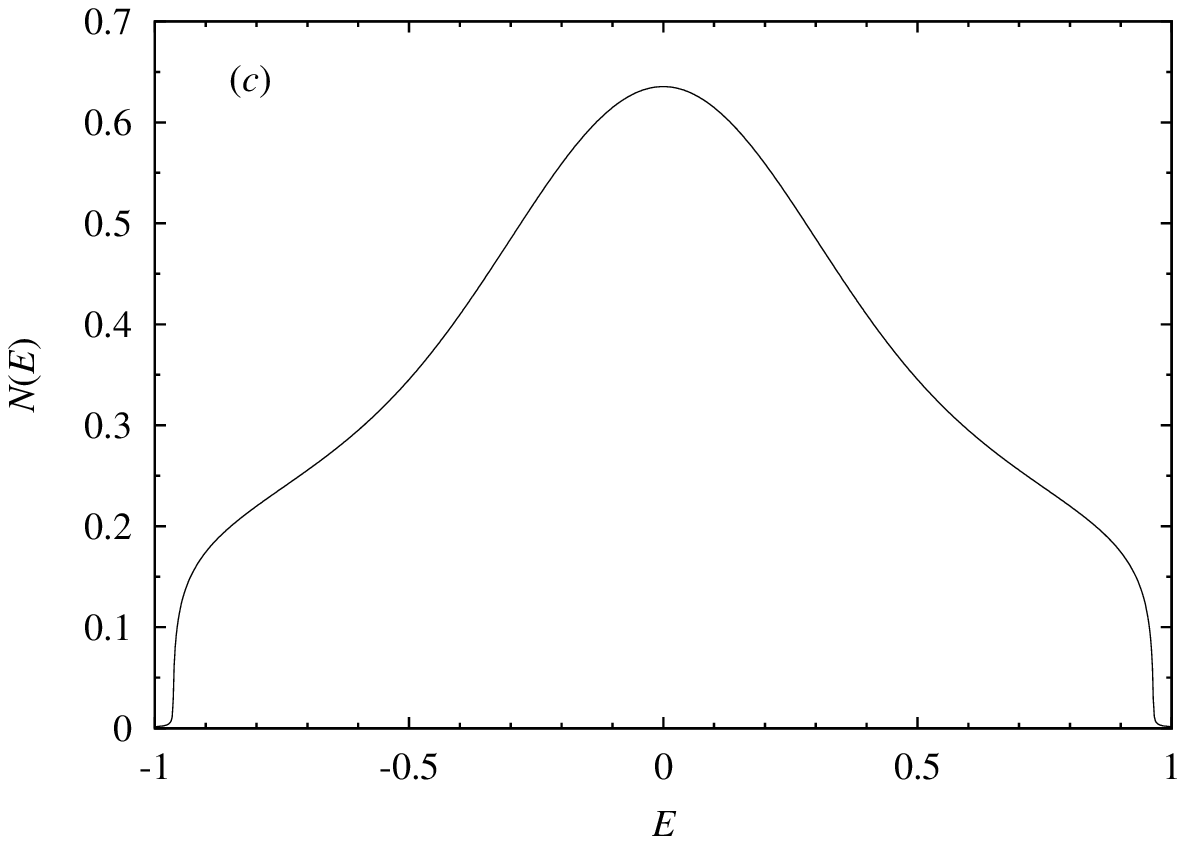}
\includegraphics[width=0.45\textwidth]{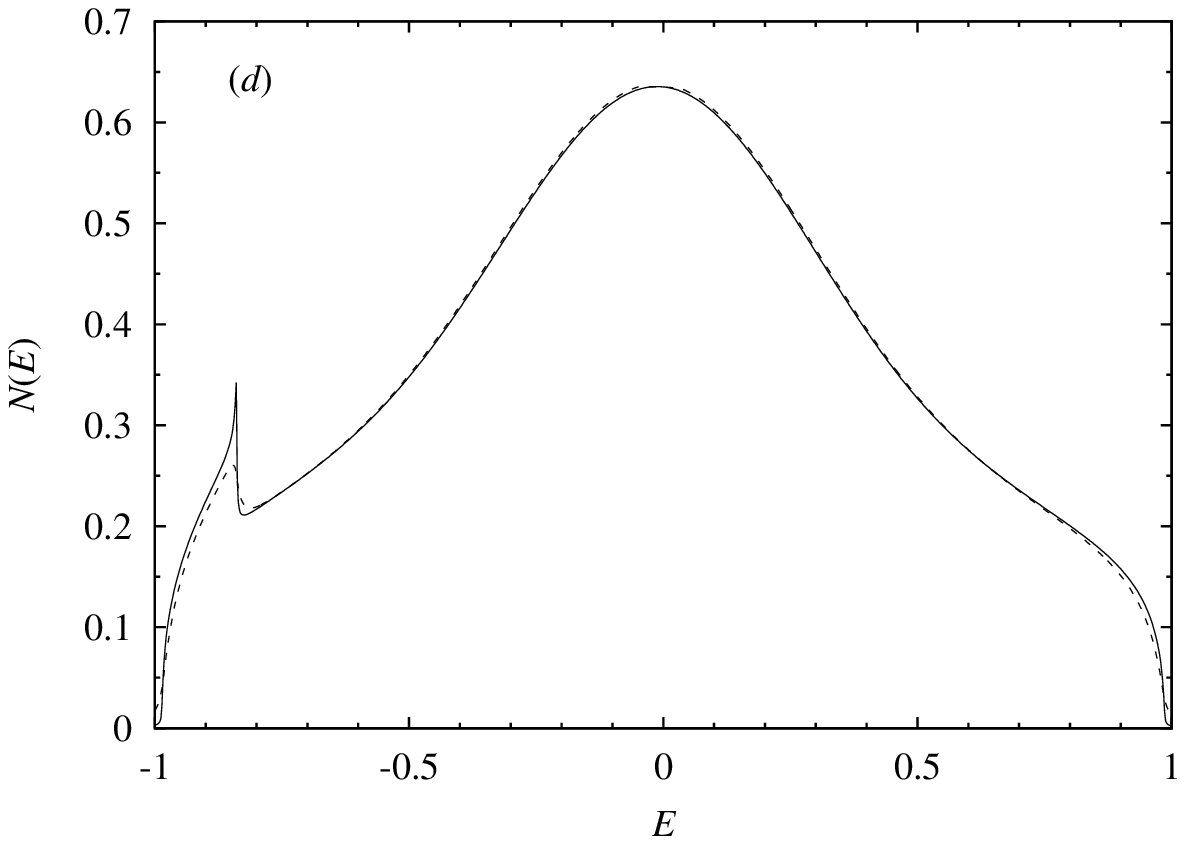}
\end{center}
\caption{Плотность состояний для полуэллиптической зоны при~$S=1/2$ и~$n=0{,}05$ в самосогласованном приближении~(\ref{eq:sd:s-c:1}). Обозначения те~же самые, что и на рис.~\ref{fig:03}}
\label{fig:04}
\end{figure}

\begin{figure}[p]
\begin{center}
\includegraphics[width=0.45\textwidth]{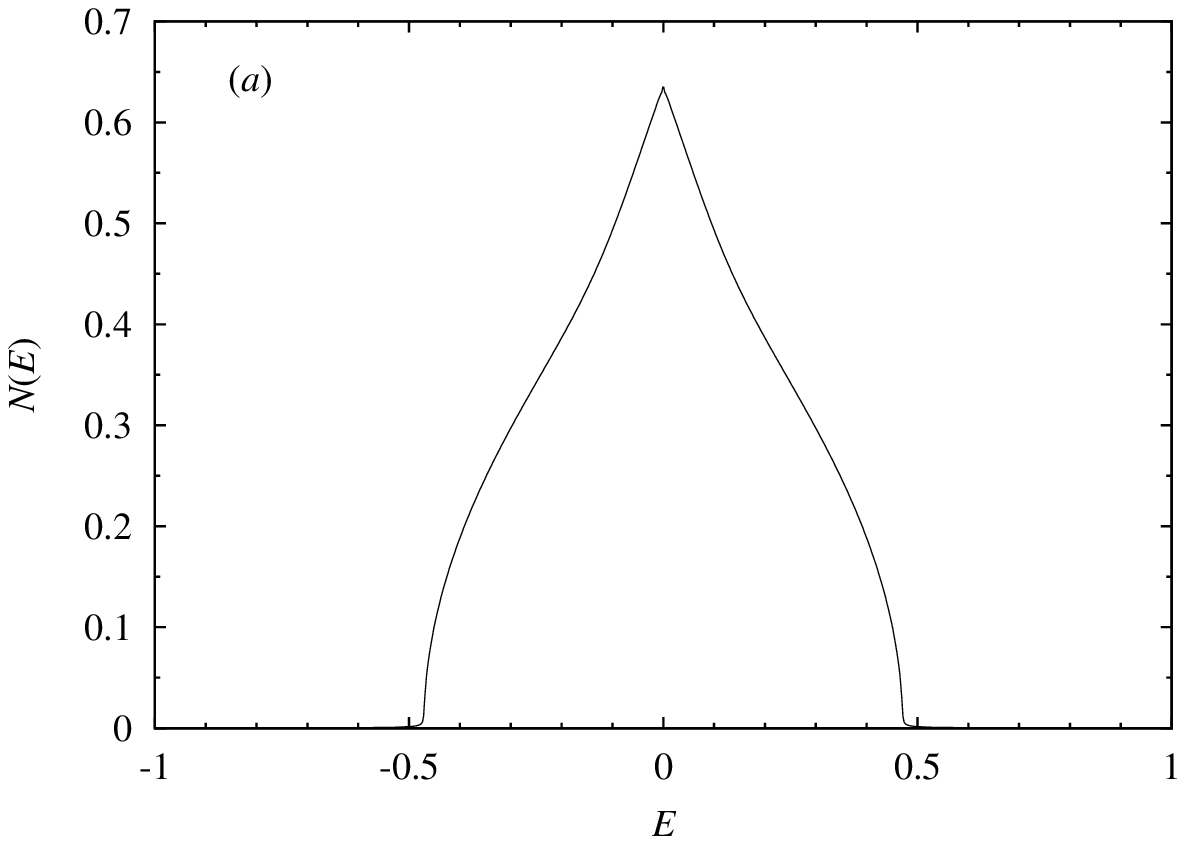}
\includegraphics[width=0.45\textwidth]{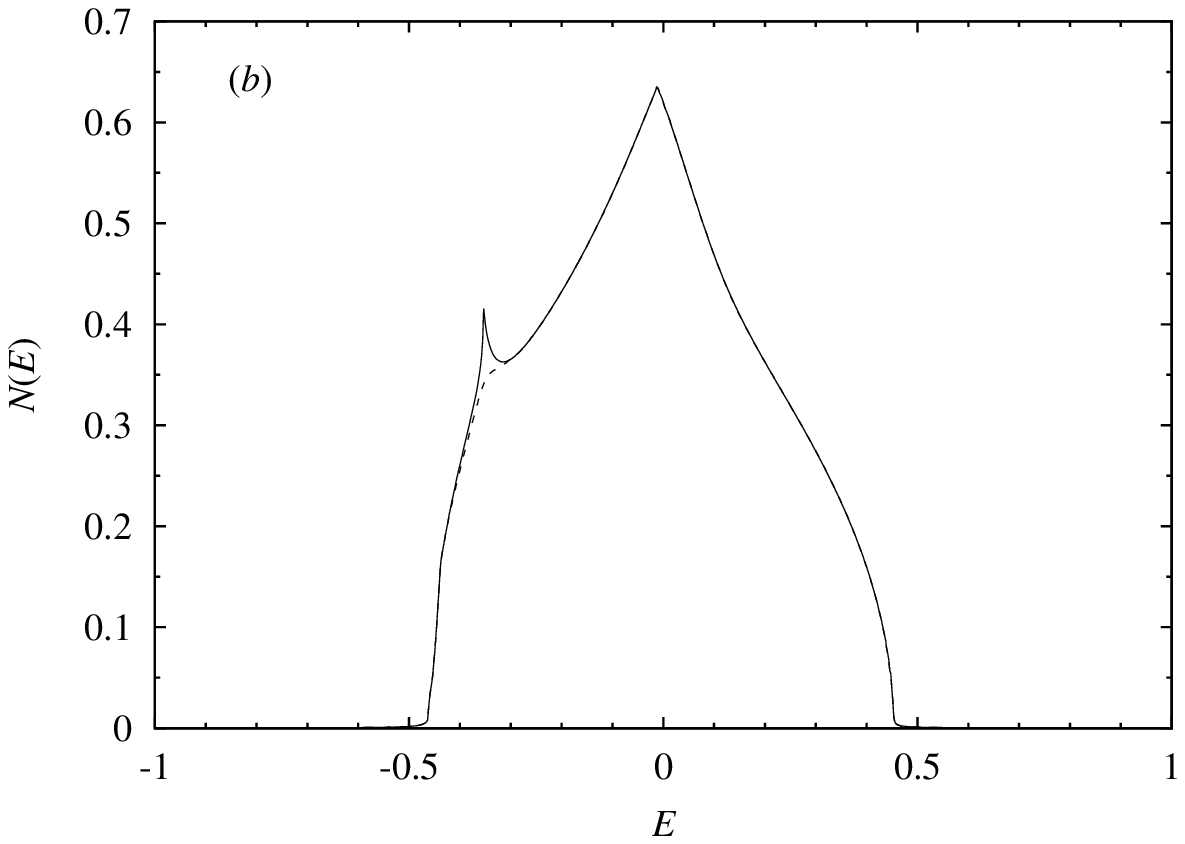}
\includegraphics[width=0.45\textwidth]{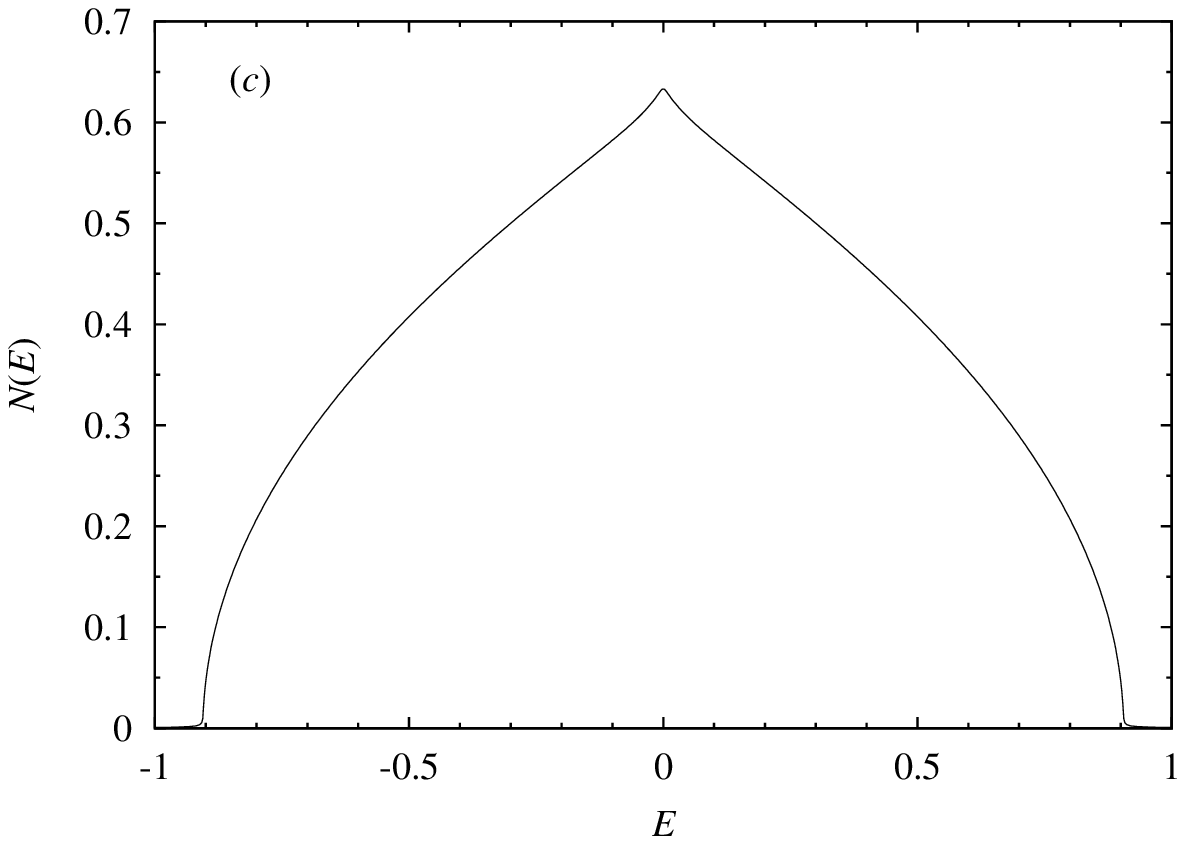}
\includegraphics[width=0.45\textwidth]{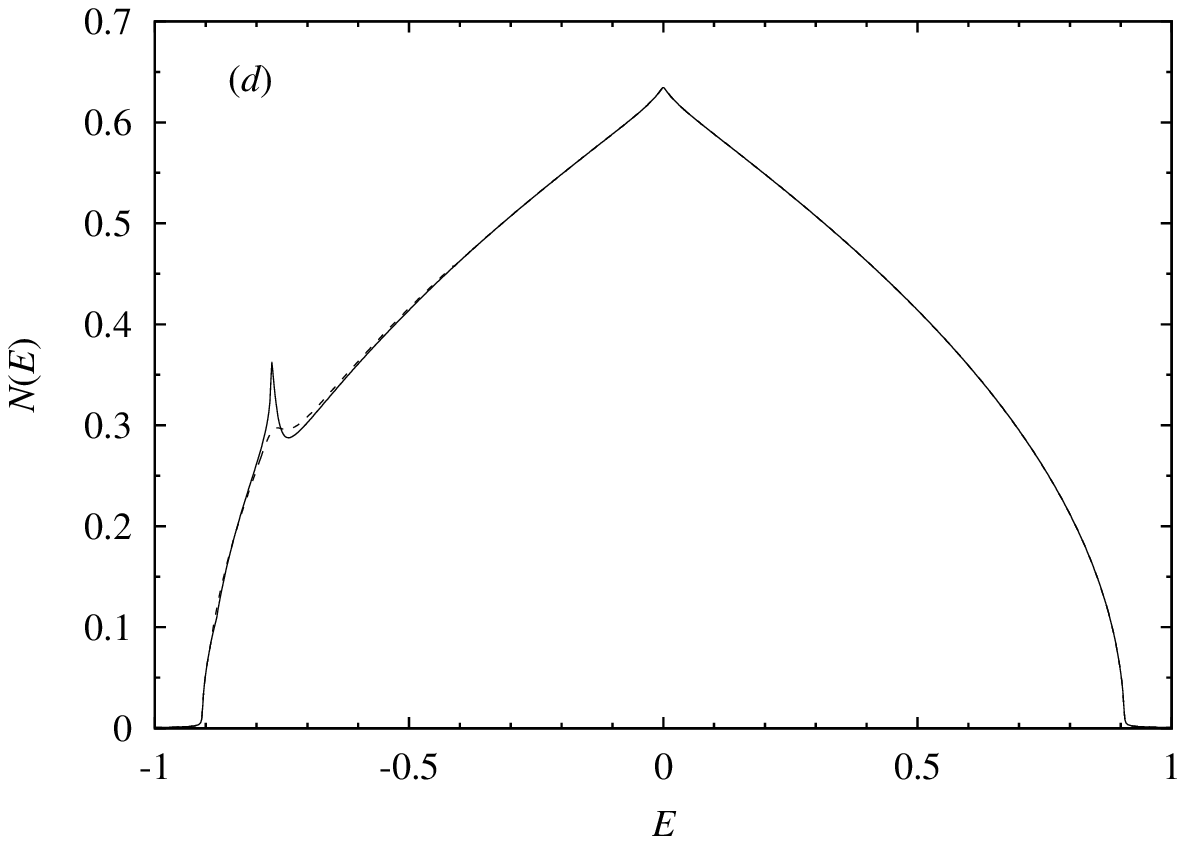}
\end{center}
\caption{Плотность состояний для полуэллиптической зоны при~$S=1/2$ и~$n=0{,}05$ в самосогласованном приближении~(\ref{eq:sd:s-c}). Обозначения те~же самые, что и на рис.~\ref{fig:03}}
\label{fig:05}
\end{figure}

Результаты численных расчётов одночастичной плотности состояний системы  в несамосогласованном приближении представлены на рис.~\ref{fig:03}, а результаты учёта самосогласований~(\ref{eq:sd:s-c:1}) и~(\ref{eq:sd:s-c}) показаны на рис.~\ref{fig:04} и~\ref{fig:05} соответственно.

Продольные и~поперечные спиновые корреляционные функции в локальном приближении переписываются в~виде
\[
\chi _{\mathbf{q}}^{\sigma -\sigma }=\chi ^{\sigma -\sigma }=\chi ,\quad \chi _{\mathbf{q}}^{zz}=\frac{1}{2}\chi _{\mathbf{q}}^{\sigma -\sigma }=\frac{1}{2}\chi ,
\]
где
\[
\chi =\left\{ \begin{array}{ll} 3(1-n)/4+2n, & \quad I\to+\infty, \\ 3(1-n)/4, & \quad I\to-\infty,\end{array}\right.
\]
$n$~"--- концентрация носителей тока.

Можно видеть, что на уровне Ферми появляется ярко выраженный пик (логарифмическая особенность)  плотности состояний.
Этот пик размывается при учёте как спиновой динамики, так и конечных температур. В наших расчетах спиновая динамика учитывалась в простейшем диффузионном приближении
\begin{eqnarray}
A_{\alpha }(E) &=&P_{\alpha }+\sum_{\mathbf{q}}\int d\omega K(\omega )\frac{t_{\mathbf{q}}}{(2S+1)^{2}}\frac{\chi +(2S+1)n_{\mathbf{q}\alpha }}{P_{\alpha }}G_{\mathbf{q}\alpha }(E-\omega ),
 \label{eq:sd:A:K} \\
B_{\alpha }(E) &=&E-\sum_{\mathbf{q}}\int d\omega K(\omega )\frac{t_{\mathbf{q}}^{2}}{2S+1}\frac{n_{\mathbf{q}\alpha }}{P_{\alpha }}G_{\mathbf{q}\alpha }(E-\omega ).
 \label{eq:sd:B:K}
\end{eqnarray}
где
\[
K(\omega )=\sum_{\mathbf{q}} K_{\mathbf{q}}(\omega),
K_{\mathbf{q}}(\omega )=\frac 1\pi \frac{\mathcal{D}q^2}{\omega ^2+(\mathcal{D}q^2)^2}
\]
$\mathcal{D}$~"--- константа спиновой диффузии.

В~самосогласованных вариантах расчёта  плотность состояний становится несколько сглаженной даже в отсутствие спиновой динамики. Следует отметить, что переход к пределу нулевой температуры не вызывает никаких трудностей, в отличие от предела больших~$d$~\cite{RMP.68.13}.

Отметим, что в DMFT (при точном решении примесной задачи) одноузельная динамика учитывается точно. Как показывают расчеты \cite{RMP.68.13}, затухание в DMFT достаточно мало, так что кондовский пик размывается слабее; особенно это заметно в центре зоны. Строго говоря, наши приближения не являются последовательно одноузельными, поскольку функциональная зависимость обратного локатора от точной одноузельной функции Грина отличается от рассмотренной в разделе~\ref{sec:dmft}.

Использование рассмотренных приближений приводит к~некоторым трудностям, которые связаны с~возникновением дополнительной особенности а в верхней полуплоскости (плоскости комплексной энергии) в приближенном выражении для функции Грина. В несамосогласованном приближении~"--- это дополнительный полюс, а в самосогласованном случае~"--- разрез.
Строго говоря, из-за этой особенности такое приближение нельзя считать "<физическим"> приближением для запаздывающей функции Грина.
Появление указанной особенности есть нарушение аналитических свойств для запаздывающей функции Грина и может привести к различного рода нарушениям, в~частности, условия нормировки.
Излом в центре зоны, видный на рис.~\ref{fig:05}, связан с обсуждавшейся неаналитичностью.
При увеличении концентрации носителей пик размывается в связи с ростом затухания, которое в нашем приближении велико; кроме того, в центре зоны начинают влиять  вклады, обусловленные неаналитичностью. Таким образом, пик размывается даже в пренебрежении спиновой динамикой.

Аналитические свойства несамосогласованного приближения для функции Грина для~положительного и~отрицательного~$I$ различны. Для~$I>0 $ ложная особенность находится в~верхней полуплоскости, а~для~$I<0$~"--- в~нижней.

Нормировка плотности состояний определяется из условия
\begin{equation}
L=\langle \{g_{i\sigma \alpha },g_{i\sigma \alpha }^{\dagger }\}\rangle =\int\limits_{-\infty }^{+\infty }N_\alpha (E)\,dE=P_\alpha .
\label{eq:sd:L}
\end{equation}

\begin{table}[p]
\caption{Величины коэффициентов нормировки~$L/P_\alpha $~(\ref{eq:sd:L}) для полуэллиптической затравочной плотности состояний}
\label{tab:2.1}
\begin{center}
\small
\begin{tabular}{|cc|cccc|}
\hline
$\alpha $    & $n$     & \multicolumn{4}{c|}{$L/P_\alpha $}    \\ \cline{3-6}
             &         & I       & II      & III     & IV      \\ \hline
$+$          & 0{,}20  & 1{,}000 & 1{,}000 & 1{,}022 & 1{,}008 \\
$+$          & 0{,}15  & 1{,}000 & 1{,}000 & 1{,}017 & 1{,}007 \\
$+$          & 0{,}10  & 1{,}000 & 1{,}000 & 1{,}011 & 1{,}005 \\
$+$          & 0{,}05  & 1{,}000 & 1{,}000 & 1{,}004 & 1{,}000 \\
$+$          & 0{,}02  & 1{,}000 & 1{,}000 & 1{,}002 & 1{,}000 \\
$+$          & 0{,}00  & 1{,}000 & 1{,}000 & 1{,}001 & 1{,}000 \\
$-$          & 0{,}00  & 1{,}510 & 1{,}395 & 1{,}325 & 1{,}014 \\
$-$          & 0{,}02  & 1{,}489 & 1{,}366 & 1{,}328 & 1{,}018 \\
$-$          & 0{,}05  & 1{,}448 & 1{,}337 & 1{,}318 & 1{,}016 \\
$-$          & 0{,}10  & 1{,}385 & 1{,}331 & 1{,}277 & 1{,}012 \\
$-$          & 0{,}15  & 1{,}342 & 1{,}246 & 1{,}227 & 1{,}010 \\
$-$          & 0{,}20  & 1{,}289 & 1{,}192 & 1{,}177 & 1{,}009 \\ \hline
\end{tabular}
\normalsize
\end{center}
I~"--- несамосогласованное приближение~(\ref{eq:sd:EGF:H1P}); II~"--- самосогласованное приближение~(\ref{eq:sd:s-c:1}); III~"--- самосогласованное приближение~(\ref{eq:sd:s-c:2}), IV~"--- самосогласованное приближение~(\ref{eq:sd:s-c}).
\end{table}

Из таблицы \ref{tab:2.1} видно, что в несамосогласованном случае  условие нормировки нарушается только для~$I<0$.
В самосогласованных приближениях (кроме~(\ref{eq:sd:s-c:1})) нарушение нормировки есть для обоих знаков $I$,
однако в приближении~(\ref{eq:sd:s-c}) оно оказывается численно малым.

Следует подчеркнуть, что проблема соблюдения правила сумм имеет общий характер и~типична для~большинства вычислений, использующих многоэлектронное представление Хаббарда или~связанные с~ним представления вспомогательных (slave, auxiliary) бозонов и~фермионов, которые включают дополнительные кинематические условия связи.
Это связано в основном с двумя причинами.
Во-первых, структура поправки теории возмущений сильной связи имеет вид $A_{\mathbf{k}}(E)t_{\mathbf{k}}$, где $A_{\mathbf{k}}(E)$  в ведущем порядке по $1/z$ или $1/d$ зависит только от энергии (это же имеет место и в локальном приближении). Поскольку зонная энергия~$t_{\mathbf{k}}$ меняет свой знак в~зоне Бриллюэна, а мнимая часть $A_(E)$, вообще говоря, отлична от нуля, то в целом поправка не имеет определенного знака мнимой части, что и приводит к появлению особенности в верхней комплексной полуплоскости. В этом случае приближение для запаздывающей функции Грина не удовлетворяет требуемым условиям аналитичности: нарушается принцип причинности и нормировка. При этом проблемы усложняются при~использовании  процедур самосогласования. В~частности, подобные трудности должны возникнуть в~приближении непересекающихся диаграмм для~модели Андерсона, где~строится разложение по~гибридизации~$V_{\mathbf{k}}$. К~сожалению, эта проблема обычно игнорируется, так~как условие нормировки фактически никогда не~проверяется.
Во-вторых, отметим, что расцепление цепочки уравнений движения в общем случае приводит к нарушению так называемых кинематических соотношений для $X$-операторов Хаббарда на одном узле. Такое нарушение вполне аналогично многократному (и, следовательно, некорректному) учету рассеяния на одном узле в теории неупорядоченных систем, который приводит к потере причинности и неправильным аналитическим свойствам в ряде случаев. Обычно выполнение кинематических соотношений в различных приближениях так же не проверяется.

\section{Заключение}

Рассмотренная в данной работе простая эффективная схема расцепления уравнений движения для функций Грина в многоэллектронном представлении позволяет воспроизвести нетривиальную структуру спектра в~случае половинного заполнения в~модели Хаббарда и воспроизвести особенности в одночастичной плотности состояний, обусловленные эффектом Кондо так же и в $s$"~$d$~обменной модели Шубина"--~Вонсовского. Рассмотренный подход даёт хорошее согласие с~результатами подходов в~случае больших размерностей пространства~$d$ и~вычислений квантовым методом Монте-Карло. В~то~же время, эти расчёты можно легко воспроизвести для~произвольной двух- и~трёхмерной решёток и обобщить на многозонный случай, что позволяет использовать подобный подход для построения сольвера в подходе DMFT. В~целом, метод многоэлектронных операторов позволяет рассмотреть регулярным способом проблему электронной структуры систем с~хаббардовским отталкиванием и с обменным $s$"~$d$~взаимодействием.

Для модели Хаббарда проанализировано влияние кулоновского взаимодействия на~эволюцию структуры плотности состояний. Построены  плотности состояний при~различных параметрах кулоновского отталкивания для~ряда затравочных модельных плотностей состояний. Показана возможность описания трёхпиковой структуры спектра, которая связана с~формированием состояний типа~"<Кондо"> в~рассматриваемой системе, а~также образование псевдощелевого состояния в~спектре вблизи перехода металл"--~изолятор при~увеличении параметра хаббардовского отталкивания электронов на~узле.

К особенностям подхода следует отнести определенную сложность построения приближения, удовлетворяющего правилу сумм и правильно учитывающего кинематические соотношения для многоэллектронных операторов. Однако в ряде случаев указанную проблему удается обойти с использованием локаторного представления для одночаcтичной функции Грина.

Работа частично поддержана программой Президиума РАН "<Квантовая физика конденсированного состояния"> и проектами РФФИ No.~11-02-00931-a и~11-02-00937-a.

\end{document}